\documentclass[aip,%
               apl,%
			   amsmath,%
			   amssymb,%
			   preprint,%
			   floatfix,
			   longbibliography]{revtex4-1}

\usepackage[separate-uncertainty,%
		    range-phrase=\text{ -- },%
            exponent-product=\cdot]{siunitx}
\usepackage{graphicx,nameref,wasysym,bm,enumitem,todonotes,setspace}
\usepackage[capitalize]{cleveref}
\usepackage[english=american,autostyle=true]{csquotes}

\usepackage{enumitem,ifthen,color}
\definecolor{grey}{rgb}{0.5, 0.5, 0.5}
\definecolor{hzbblue}{rgb}{0., 0.34375, 0.609375}

\newcommand{\dif}{\,\mathrm{d}}

\renewcommand{\vec}[1]{\bm{\mathbf{#1}}}

\newcommand{\mat}[1]{\widetilde{#1}}

\newcommand{\abs}[1]{\left|{#1}\right|}
\newcommand{\norm}[1]{\left\|{#1}\right\|}

\newboolean{preprint}
\setboolean{preprint}{false}
\newcommand{\condprint}[1]{\ifthenelse{\boolean{preprint}}{#1}{}}

\newcommand{\omitln}[1]{\ifthenelse{\isundefined{\linenumbers}}{#1}{\nolinenumbers

#1

\linenumbers}}
\newcommand{\forceln}[1]{\ifthenelse{\isundefined{\linenumbers}}{#1}{\begin{linenumbers}
		
		#1
		
		\end{linenumbers}}}

\begin{document}

\author{Carlo Barth}
\author{Christiane Becker}
\email{christiane.becker@helmholtz-berlin.de}
\affiliation{Helmholtz-Zentrum Berlin f\"{u}r Materialien und Energie, %
	Kekul\'{e}str.~5, \mbox{12489 Berlin,~Germany}}

%%%%%%%%%%%%%%%%%%%%%%%%%%%%%%%%%%%%%%%%%%%%%%%%%%%%%%%%%%%%%%%%%%%%%
%\title{Machine learning for photonic crystals: classification of leaky modes using clustering}
%\title{Machine learning classification of photonic crystal leaky modes}
%\title{Machine learning classification of photonic mode field distributions}
\title{Machine learning classification for field distributions of photonic modes}


% Date
\date{\today}

%%%%%%%%%%%%%%%%%%%%%%%%%%%%%%%%%%%%%%%%%%%%%%%%%%%%%%%%%%%%%%%%%%%%%
% ~ 150 words
\begin{abstract}
	Machine learning techniques can reveal hidden structure in large data amounts and can potentially extent or even replace analytical scientific methods. In nanophotonics, modes can increase the light yield from emitters located inside the nanostructure or near the surface. Optimizing such systems enforces to systematically analyze large amounts of three-dimensional field distribution data. We present a method based on finite element simulations and machine learning for the identification of modes with large field energies and specific spatial properties. By clustering we reduce the field distribution data to a minimal subset of prototypes. The predictive power of the approach is demonstrated using an analysis of experimentally measured fluorescence enhancement of quantum dots on a photonic crystal surface. The clustering method can be used for any optimization task that depends on three-dimensional field data, and is therefore relevant for biosensing, quantum dot solar cells or photon upconversion.
%\forceln{ \struc{background:}. \struc{rationale:} .\struc{main results:} . \struc{implications:} .} \note{Word count: \textbf{0}}
\end{abstract}

% Title
\omitln{\maketitle}

%%%%%%%%%%%%%%%%%%%%%%%%%%%%%%%%%%%%%%%%%%%%%%%%%%%%%%%%%%%%%%%%%%%%%
%% Start the main part of the manuscript here.
%%%%%%%%%%%%%%%%%%%%%%%%%%%%%%%%%%%%%%%%%%%%%%%%%%%%%%%%%%%%%%%%%%%%%

\noindent Machine learning is a rapidly developing discipline, which uses statistical approaches to learn from data without explicitly rule-based programming. Driven by today's massive increase in data amounts, the related techniques are extended and improved at a fast pace\cite{Jordan2015}. Machine learning is currently applied to all aspects of science, from health sciences and psychology\cite{Just2017,Guncar2018}, to biology\cite{Steinegger2018,Chen2018,Kan2017}, to environmental\cite{Exbrayat2017} and material sciences\cite{Sumpter2015,Ramprasad2017}. But also to matters of everyday-life, from online security, to finance and insurance. While supervised learning has led to breakthroughs in computer vision\cite{Russakovsky2015} and speech recognition\cite{Hinton2012}, unsupervised learning is expected to become far more important in the future\cite{Lecun2015}. The latter techniques, such as clustering\cite{Jain1999,Xu2005,Aghabozorgi2015}, allow for the recognition of patterns in unlabeled data and can therefore reveal a hidden structure. They have been successfully applied to e.g.~anomaly detection\cite{Garcia-Teodoro2009,Pimentel2014} or genetics\cite{Libbrecht2015}.

In the field of nanophotonics, increasing computer power, storage space and data throughput, as well as improvements in modeling techniques, greatly accelerated all-numerical system design. For nanostructures the following typical optimization tasks are met:
\begin{itemize}
	\item \textbf{Simple design:}\newline
	Scalar parameters $\rightarrow$ scalar output\newline
	(e.g.~lengths/refractive indices $\rightarrow$ reflectivity)
	\item \textbf{Inverse design:}\newline
	Multivariate parameters $\rightarrow$ scalar output\newline
	(e.g.~permittivity distribution $\rightarrow$ reflectivity)
	\item \textbf{Qualitative design:}\newline
	Scalar parameters $\rightarrow$ multivariate output\newline
	(e.g.~lengths/refractive indices $\rightarrow$ 3D field distribution)
\end{itemize}
Simple design tasks are generally solved by simulating the system for many different parameter combinations (i.e.~grid search), or by applying function minimization routines. More sophisticated techniques such as the reduced basis method\cite{Hammerschmidt2016} for finite element method (FEM) simulations have successfully been applied to speed-up this optimization process for large parameter spaces. Inverse design tasks introduce a high-dimensional input parameter space, typically by allowing for arbitrary changes in the permittivity distribution $\epsilon(\vec r)$ of the nanostructure. Machine learning techniques have successfully been applied for this purpose, mainly using genetic algorithms\cite{Smajic2004,Hakansson2005,Lu2013,Lu2013a,Piggott2015}. Simple and inverse tasks have in common that they possess a scalar measure of success, i.e.~they can be seen as minimization problems. The machine learning approach in inverse design therefore belongs to the field of supervised learning (more specifically regression). The third design task introduced above substantially differs in the way that the system should be optimized for a high-dimensional \emph{output}. Due to the inaccessibility of a scalar success metric, we denote this problem as qualitative design. This is for example the case if the 3-dimensional spatial distribution of the electromagnetic fields has to be taken into account. Usually, such problems are solved by appropriate visualizations. But since any change in the input parameters leads to a change in the high-dimensional output, the data amounts quickly become extremely large. We will demonstrate below that machine learning, or more specifically clustering, is able to overcome these issues by reducing the output dimensionality.

As indicated before, an example of qualitative design is to optimize a photonic nanostructure, e.g.~a photonic crystal (PhC), for an appropriate spatial field distribution. This is of high relevance whenever an interaction of the field with a (potentially vague) particle distribution is present, e.g.~for emitters on nanophotonic surfaces or emitters embedded into the nanostructure. PhC slabs exhibit a phenomenon called \emph{leaky modes}: resonances that can be excited using external radiation\cite{Rosenblatt1997,Astratov1999,Astratov1999a,Erchak2001,Ochiai2001}. Leaky modes have been used to improve various applications (e.g.~light trapping in photovoltaic devices\cite{Chutinan2008,Han2010a,John2012,Mellor2013,Branham2015}, light-emitting diodes \cite{Fan1997,Wiesmann2009}), but can also affect near-surface emitters, such as QDs, atoms or molecules. Especially in the life sciences, the applications range from PhC enhanced microscopy and single molecule detection to enhanced live cell imaging, DNA sequencing and gene expression analysis\cite{Cunningham2016,Block2009,Ganesh2008a,Threm2012}. Besides the rather well-investigated extraction enhancement effect\cite{Astratov1999,Astratov1999a,Boroditsky1999,Erchak2001,Fan1997,Ganesh2008,Ondic2011a,Ondic2012,Ondic2013}, the excitation enhancement effect\cite{Adachi2013,Kim2013,Liu2013,Zhang2010a,Hofmann2016,Ganesh2007,Ganesh2008a} increases the stimulated emission rate of the emitters by enhanced near-field energy densities of leaky modes in the absorption wavelength range. To optimize photonic nanostructures for excitation enhancement it is therefore inevitable to take the 3D spatial electromagnetic field distribution into account.

\begin{figure*}[t]
	\centering \includegraphics{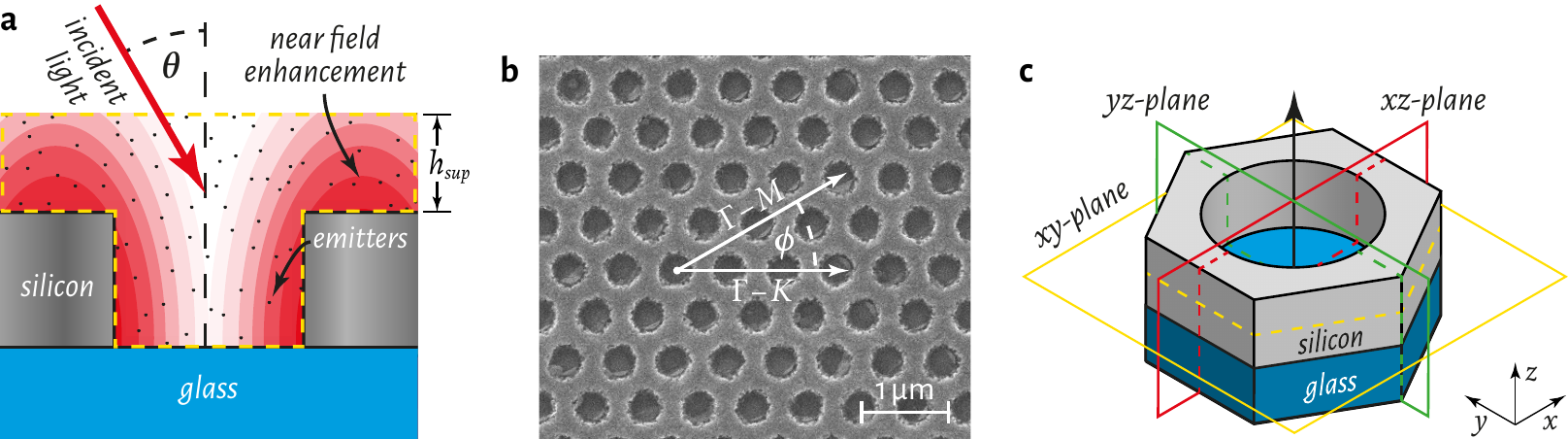}
	\caption{\textbf{Overview of the nanostructure}. (\textbf{a}) Light incident on a silicon photonic crystal (PhC, gray) on glass (cyan) excites a leaky mode that exhibits enhanced electromagnetic near-field energies in the superspace volume (marked by the yellow dashed line). Emitters (black dots) in the vicinity of the PhC surface interact with the local electric field distribution. (\textbf{b}) Scanning electron microscopy image of the PhC sample with denoted high-symmetry directions $\Gamma-K$ and $\Gamma-M$. (\textbf{c}) A unit cell of the PhC system as used in the simulation. Yellow, green and red rectangles mark the planes used for the field export.} \label{fig:overview}
\end{figure*}

In this study we present a powerful technique based on machine learning for the classification of 3D electromagnetic field distribution data. This method can be of avail in any case where large amounts of electromagnetic field (or energy) distribution data should be reduced to a minimal subset of typical distributions. We will refer to these as \emph{distribution prototypes}. We directly apply the technique to a specific dataset of our previous publication on fluorescence enhancement of lead sulfide (PbS) quantum dots (QDs) on a silicon PhC slab surface\cite{Barth2017}, however, without loss of generality. A similar setup was used in previous studies \cite{Becker2014a,Barth2016,Hammerschmidt2016}.

%PhC slabs exhibit a phenomenon called \emph{leaky modes}: resonances that can be excited using external radiation\cite{Rosenblatt1997,Astratov1999,Astratov1999a,Erchak2001,Ochiai2001}. Leaky modes have been used to improve various applications (e.g.~light trapping \cite{Chutinan2008} and light-emitting diodes \cite{Fan1997,Wiesmann2009}), but can also affect near-surface emitters, such as QDs, atoms or molecules. Especially in the life-sciences, the applications range from photonic crystal enhanced microscopy and single molecule detection to enhanced live cell imaging, DNA sequencing and gene expression analysis\cite{Cunningham2016,Block2009,Ganesh2008a,Threm2012}. Besides the rather well-investigated extraction enhancement effect\cite{Astratov1999,Astratov1999a,Boroditsky1999,Erchak2001,Fan1997,Ganesh2008,Ondic2012,Ondic2011a,Ondic2011,Ondic2013}, the excitation enhancement effect\cite{Adachi2013,Kim2013,Liu2013,Zhang2010a,Hofmann2016,Ganesh2007,Ganesh2008a} increases the stimulated emission rate of the emitters by enhanced near-field energy densities of leaky modes in the absorption wavelength range.

The effect is sketched in \cref{fig:overview}(a), depicting emitters (black dots) that interact with a leaky mode of the PhC excited by an external laser source. The illumination conditions introduce four parameters: the laser wavelength $\lambda$, the laser polarization $\mathcal{P}$ (TE or TM), the polar angle $\theta$ with the plane normal, and the azimuthal angle $\phi$ used to define the high symmetry direction ($\Gamma-M$ or $\Gamma-K$). The latter is also indicated in the scanning electron microscope image of the sample (without emitters) in \cref{fig:overview}(b). An example of an electric field distributions $\vec{E}(\vec{r})$ of a leaky mode is depicted in \cref{fig:overview}(a). As mentioned, the energy density of the electric field of the leaky modes, $w_\mathrm{lm}(\vec{r})$, can be larger compared to the energy density of the incident plane wave, $w_\mathrm{pw}$, known as field energy enhancement ($w_\mathrm{lm}(\vec{r})/w_\mathrm{pw}>1$). To study this effect in large parameter spaces we usually define the \emph{volume-integrated} field energy enhancement
\begin{equation}
E_+ = \frac{1}{w_\mathrm{pw}V_\mathrm{sup}} \int_{V_\mathrm{sup}}w_\mathrm{lm}(\vec{r})\dif{V_\mathrm{sup}},
\label{eq:e_enhancement}
\end{equation}
where $V_\mathrm{sup}$ is the volume of interest. In our case $V_\mathrm{sup}$ is the superspace of the computational domain, as indicated by the yellow dashed line in \cref{fig:overview}(a). The energy density of the plane wave has no spatial dependence and is proportional to the amplitude of the electric field, $\vec{E}_{\mathrm{pw,}0}$, and the refractive index $n$ of the surrounding medium, i.e.
\begin{equation}
w_\mathrm{pw} = \frac{\epsilon_0}{4}n^2\norm{\vec{E}_{\mathrm{pw,}0}}^2.
\end{equation}
In the figure, a uniform random distribution of emitters is shown as an example. But depending on the coating process, emitters might have a very specific spatial distribution in a real application, e.g.~a monolayer attached to the surface, or a higher concentration inside the holes, or at the plateaus between the holes. Consequently, the spatial distribution of the energy density $w_\mathrm{lm}(\vec{r})$ becomes a determining factor and, therefore, the integrated field energy enhancement $E_+$ is not sufficient to quantify the effect on the emitters.  %The leaky modes form photonic bands with spatial properties that depend on the symmetry of the system, and which are conserved quantities. The number of spatial distribution types is therefore finite and, more specifically, of the order of the number of bands in the parameter window. 

\begin{figure*}[t]
%	\centering \includegraphics{figures/enhancement_vs_clustering.pdf}
	\centering \includegraphics{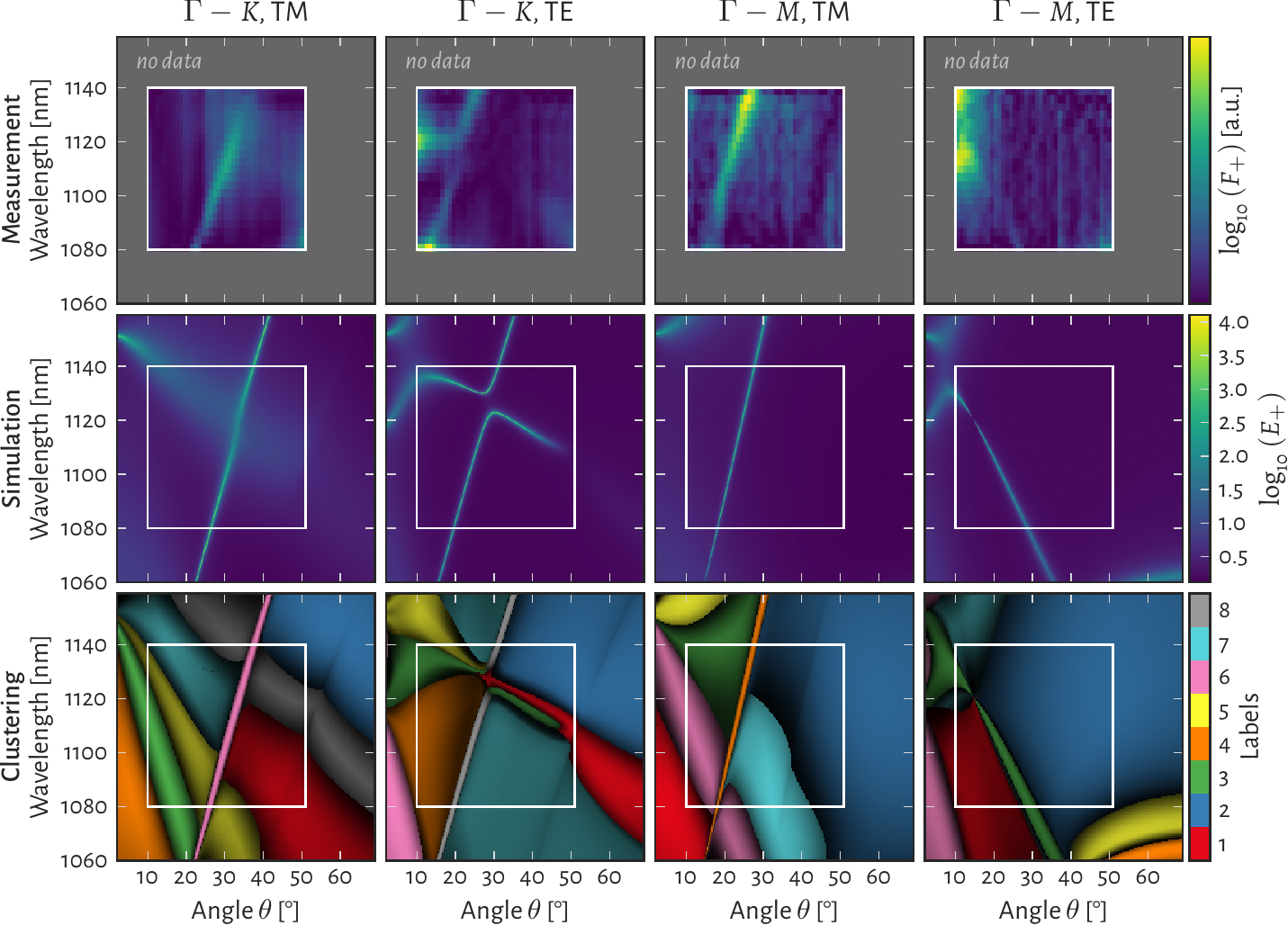}
	\caption{\textbf{Comparison of measured quantum dot fluorescence enhancement $F_+$, simulated volume-integrated field energy enhancement $E_+$, and corresponding classification maps.} (\textbf{Upper row}) Measured fluorescence enhancement $F_+$ as a function of vacuum wavelength and incident angle $\theta$ of the laser source (logarithmic color scale; see \cref{fig:overview}(a) which indicates $\theta$; see supplementary material of Ref.~\citenum{Barth2017} for experimental setup). The columns correspond to the four combinations of sample orientation ($\Gamma-M$ and $\Gamma-K$) and source polarization (TE and TM). (\textbf{Center row}) Simulated volume-integrated electric field energy enhancement $E_+$ for the same conditions as in the upper row. For a definition of the volume $V_\mathrm{sup}$ see \cref{fig:overview}(a). The white lines mark the experimental data limits. (\textbf{Lower row}) Classification maps depicting the cluster assignments (labels) using different colors independently for each plot, and the respective silhouette coefficients using alpha-blending with a black background (color bar omitted; see \cref{fig:silhouettes} for value ranges). More saturated colors denote larger silhouette coefficients. (Note: Upper and center rows of the panel repeat the same results as already shown in Ref.~\citenum{Barth2017} for a larger angle and wavelength range.)} \label{fig:enhancement_vs_clustering}
\end{figure*}

An optimized design for an application as sketched in \cref{fig:overview}(a) can hence be achieved by identifying a mode which has (i) a large volume-integrated field energy enhancement $E_+$ \emph{and} (ii) an appropriate spatial field energy density distribution overlapping the locations of the emitters, at the same time. Task (i) is a \enquote{simple design} task, as defined in the introduction, while (ii) is a \enquote{qualitative task}, i.e.~an optimization of a multivariate output. %
%In the same way, it is necessary to characterize the system in view of these two properties to explain measured fluorescence enhancement effects like the ones of our previous study\cite{Barth2017}.%
The results of the fluorescence enhancement measurements of our previous study\cite{Barth2017} are shown again in the upper row of \cref{fig:enhancement_vs_clustering}, as well as the results for the volume-integrated field energy enhancement $E_+$ in the center row. The latter results solve task (i), as described above. Task (ii) potentially enforces to take into account 3D field distribution data of \emph{all} combinations of the illumination condition parameters $\lambda$, $\mathcal{P}$, $\theta$ and $\phi$. If the number of considered wavelengths $N_\lambda$ and the number of angles $N_\theta$ becomes large, it is no longer feasible to directly visualize all the 3-dimensional field distributions for all points in the $\lambda$-$\theta$-maps shown in \cref{fig:enhancement_vs_clustering}. It is hence necessary to reduce the amount of field distribution data in an appropriate way. One possibility to achieve this reduction is to pitch on specific wavelengths and incident angles for which the field distribution is evaluated, as it was done in the previous study\cite{Barth2017}. This way, however, information is mainly gained at random, so that general trends might be overseen. A more systematical approach is to \emph{cluster} field distributions which are similar, and to therefore derive typical distributions (i.e.~\enquote{distribution fingerprints}). It is known that a certain undisturbed photonic band in the leaky mode regime will not significantly change its symmetry properties when crossing the $\lambda$-$\theta$ space\cite{Ochiai2001,Sakoda2004}, as will be explained in more detail below. As a result, the entirety of field distributions are composed of a finite set of patterns which are basically caused by the finite number of bands. This feature space can efficiently be partitioned into the typical patterns using machine learning clustering techniques.

In the following, we will first reconsider the experimental and numerical results of the previous study\cite{Barth2017}, highlighting aspects which were left unexplained by the prior analysis technique. Afterwards, we introduce the clustering technique and apply it to systematically analyze the 3D field energy distribution properties. The distributions are classified by assigning them to distribution prototypes, which are consulted as representative solutions to fully explain the effects observed in the experiment. We further consider a mathematical method based on silhouette coefficients\cite{Rousseeuw1987} to assess the clustering result. Based on these analyses we will explain how the method enables to solve complex optimization tasks with high-dimensional output, as indicated in the introduction. The all-numerical technique is of relevance for the design of nanophotonic structures for any application in which emitters interact with the electromagnetic field, e.g.~highly-sensitive biosensors\cite{Cunningham2016}, quantum dot solar cells\cite{Carey2015a,Kamat2008a,AkihiroKojimaKenjiroTeshimaYasuoShirai2009,Kundu2017}, or up-conversion devices\cite{Wu2015b,Wu2015c}.

% =============================================================================
\section*{Results}\label{sec:results}

The upper two rows of \cref{fig:enhancement_vs_clustering} repeat the main findings of the prior fluorescence enhancement study\cite{Barth2017}. The upper row shows the fluorescence enhancement ($F_+$) maps obtained by tilting the QD-coated PhC sample along the respective high-symmetry directions of the irreducible Brillouin zone ($\Gamma-M$ or $\Gamma-K$, adjusted using $\phi$), and by using transverse-electric (TE) or transverse-magnetic (TM) polarization $\mathcal{P}$ of the incident laser radiation. Each measured spectrum (for a single incident angle) was first integrated over the fluorescence peak from $\lambda=\SI{1200}{\nm}$ to $\lambda=\SI{1700}{\nm}$ and normalized to the measured incident laser power and the absorption profile of the QDs, yielding the fluorescence $F$. A minimum estimate for the fluorescence enhancement $F_+$ is obtained from dividing by the minimal value in each of the maps. The maps feature regions of enhanced fluorescence.

The measured fluorescence enhancement is caused by increased energy densities of the fields at the emitter positions. Concerning task (i) of the introduction, the center row of \cref{fig:enhancement_vs_clustering} maps the electric field energy enhancement $E_+$ integrated over the simulated superspace volume $V_\mathrm{sup}$, which contains the QDs (see \cref{eq:e_enhancement}). The $E_+$ maps exhibit clearly visible bands of strong field energy enhancement, which partly correspond to regions of high measured fluorescence $F_+$. Some deviations are caused by a Q-factor mismatch between the spectral bandwidths of the leaky modes and the excitation laser source. However, a few features of the measured $F_+$ maps remained unexplained, for example:
\begin{itemize}
%	\item $\Gamma-K$, TM: the region of high fluorescence enhancement at $\sim(\ang{55}, \SIrange{1080}{1120}{\nm})$,
	\item $\Gamma-K$, TE: The declining band after the anticrossing point, which is visible in the corresponding $E_+$-map, but missing in the experimental fluorescence enhancement $F_+$.
	\item $\Gamma-M$, TE: The elongated bright spot of high fluorescence enhancement at about $(\ang{10}, \SIrange{1100}{1140}{\nm})$.
\end{itemize}
Please consult Ref.~\citenum{Barth2017} for further details of the comparison.

\subsection*{Introduction and justification of the clustering technique}

The $E_+$-maps given in the center row of \cref{fig:enhancement_vs_clustering} only provide information about the \emph{volume-integrated} field enhancement over a characteristic volume $V_\mathrm{sup}$, marked by the yellow dashed line in \cref{fig:overview}(a). Therefore, regions of high $E_+$ can be regarded as a necessary condition for fluorescence enhancement, but not as a sufficient one. A high $E_+$ without a corresponding fluorescence enhancement $F_+$ hence indicates a lack in the spatial overlap of the emitters with the regions of enhanced field energy density.

However, it is known that bands of the photonic crystal have well-behaved spatial properties when varying the $k$-vector between two high-symmetry points of the irreducible Brillouin zone\cite{Ochiai2001,Sakoda2004}. More specifically, the modes belong to the same symmetry point group as the system seen from the point in $k$-space, i.e.~they exhibit the same spatial symmetry. Consequently, it is theoretically justified to expect that the spatial properties of the bands only change smoothly with $\theta$. For a fixed high-symmetry direction, e.g.~$\Gamma-K$, we only expect two types of solutions, which are
\begin{enumerate}
	\item regions that correspond to leaky-mode bands, and
	\item regions that are off any photonic band, and therefore corresponding to the continuum of radiation modes.
\end{enumerate}
The regions of the radiation modes are expected to exhibit solutions that resemble plane waves, i.e.~show oscillatory behavior in the the exterior domain. All things considered, only a small number of different spatial symmetry types is expected, which is of the order of the number of bands that cross the parameter scan window.

\begin{figure*}[t]
	\centering \includegraphics{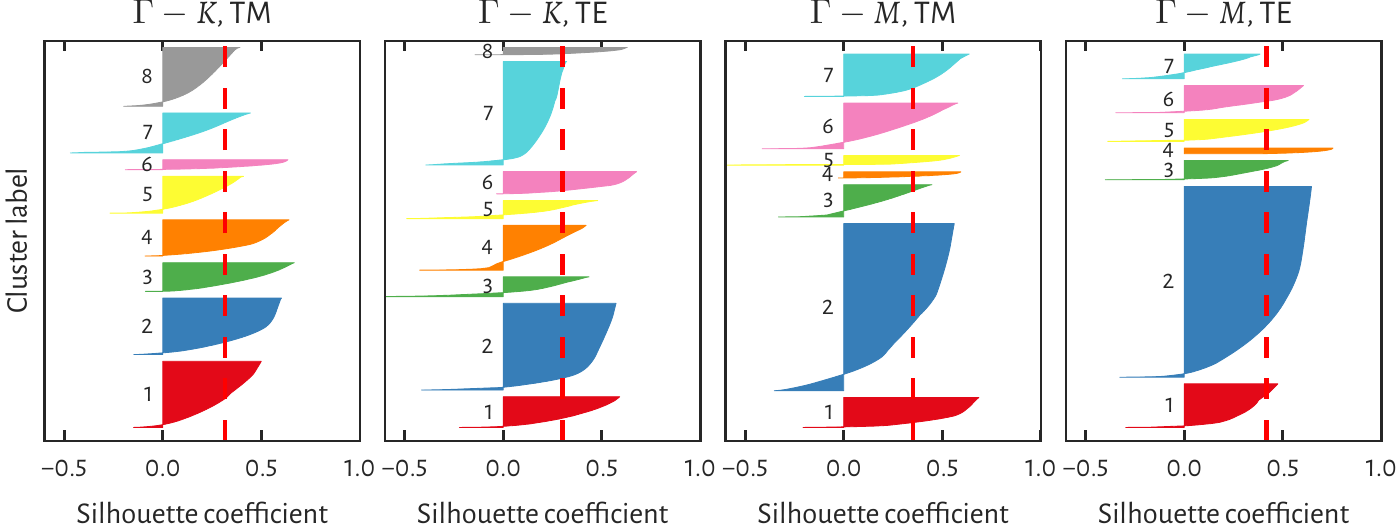}
	\caption{\textbf{Silhouette analysis plots for the different direction/polarization combinations.} In each of the silhouette plots (i.e.~columns), silhouette coefficients for each sample are plotted as a bar in $x$-direction with a length corresponding to its value. The samples are sorted by their silhouette coefficients, with smaller values being located at smaller $y$-positions; and grouped and color-coded using the same colors as in \cref{fig:enhancement_vs_clustering}. Red dashed lines mark the silhouette scores.} \label{fig:silhouettes}
\end{figure*}

This is where machine learning comes into play. If we consider the specific 3D electric field distribution for a single illumination setting as a \emph{sample}, and the electric field values of each point in the considered volume as \emph{features}, then clustering techniques are able to subdivide the entirety of field distributions into a finite number of field distribution prototypes. This approach is reasonable because the data range is expected to contain a finite number of typical field patterns, and each of the real field patterns can be identified with one of those prototypes. Moreover, these prototypes have a sufficient \enquote{uniqueness}, e.g.~they considerably differ in their symmetry properties. The methods section \enquote{Clustering of electric field data} gives a detailed description of how the clustering is performed. In a nutshell, for each illumination condition, i.e.~a set of $(\mathcal{P}, \phi, \theta, \lambda)$, the electric field strength $E=(E_x,E_y,E_z)$ is derived from a FEM simulation. It is sufficient to export the fields on symmetry planes to reduce the data volume, for which we use the $xy$, $xz$ and $yz$ planes marked in \cref{fig:overview}(c). The validity of this approach was tested using a comparison to full 3D exports using a smaller dataset. Note that, in contrast to the the volume-integrated field energy enhancement $E_+$ which is calculated in the volume $V_\mathrm{sup}$ (\cref{fig:overview}(a)), the fields for the clustering are also considered in the dielectric materials (silicon PhC and glass substrate, \cref{fig:overview}(c)). To account for the different cluster sizes (narrow bands) and unknown cluster shapes in the data set, the flexible Gaussian mixture model (GMM) clustering technique is used (see the methods section \enquote{Gaussian mixture model clustering} for details), implemented in the Python library Sckit-learn\cite{Pedregosa2012}. From the clustering itself two characteristics can directly be gained: the classification, which \emph{labels} each observation with a cluster index $i$, and the \emph{distribution prototypes}, usually denoted as \enquote{cluster centers} in the general clustering literature.

The latter are the average of the electric field distributions (on the chosen planes) of all samples that belong to a specific cluster $i$. We note that we averaged the \emph{normalized} input data, i.e.~the exact data used for the clustering, to calculate the prototypes. The prototypes therefore represent the actual mathematical cluster centers, with the tradeoff that the absolute field amplitude information is lost, because the samples are normalized individually. Another possibility would be to average the unnormalized fields, so that the amplitude information would be conserved, with the tradeoff that the prototypes derived that way are not exactly the cluster centers. We settled for the normalized fields, as the amplitude information is essentially included in the $E_+$ maps.

As in most clustering techniques, the number of clusters must be specified in GMM clustering, so that the appropriateness of this choice has to be validated. This aspect will be covered shortly.

\subsection*{Classification maps}

The classification can be visualized by assigning each point $(\theta_j,\lambda_k)$ to a different color that corresponds to its label $i$. Recall that the clustering is carried out individually for each combination of polarization $\mathcal{P}$ and azimuthal angle $\phi$ ($=$high symmetry direction $\Gamma-M$ or $\Gamma-K$). Plotted in the same fashion as the $E_+$ maps of the center row of \cref{fig:enhancement_vs_clustering}, we denote the resulting figures as \emph{classification maps}. These classification maps are shown in the lower row of the figure. The color scale relates the colors to the labels and, hence, identify the corresponding cluster. Note that the classification maps cannot be compared among each other, although the same colors have been used. The clusterings for the $\Gamma-K$-cases used 8 clusters, while the $\Gamma-M$-cases only required 7 (i.e.~there is no gray region in these maps). The procedure of determining the number of clusters will be explained in the next subsection.

When comparing the classification maps to the $E_+$ maps above, a striking accordance can be observed. The narrow bands of high field enhancement in the $E_+$ maps correspond to narrow areas at the same positions in the classification maps. Note that the $E_+$ maps and the classification maps are based on very different data sets: the former are derived from a spatial integration over the electric field energy density distribution $w_\mathrm{lm}(\vec{r})$ in the superspace volume $V_\mathrm{sub}$ only (\cref{eq:e_enhancement}), while the latter uses electric field patterns $\vec{E}(\vec{r})$ on planes that include the PhC and glass domains. When observing the regions off the leaky-mode bands, i.e.~the domains of the radiation modes, it is seen that these regions are multiply subdivided in some cases; e.g.~$\Gamma-K$, TM: bottom left. In contrast, other parts are homogenous over large ranges, such as $\Gamma-M$, TE: top right.

Another detail of these plots are the different levels of saturation used for each point, obtained by alpha-blending with a black background. This additional layer of information illustrates the \emph{representation quality} of the local solution by the assigned cluster, as determined using so-called silhouette coefficients\cite{Rousseeuw1987}. The silhouette coefficients provide a way to assess the initial choice of the number of clusters, \emph{and} how well the samples lie in their respective clusters, at the same time. The silhouette coefficient rates how well a sample fits into its own cluster. If it is far away from all other clusters and very close to the cluster center (i.e.~prototype), the sample gets a positive rating. If the distances to a different cluster and its own cluster are comparable, it is rated with values close to zero. Finally, if it is much closer to a different cluster, a negative rating is assigned. See the methods section \enquote{Solution quality rating using silhouette coefficients} for a severe definition.

In all cases, we observe that the saturation decreases at the border of two clusters. This is expected, as silhouette scores close to 0 indicate a sample which is in fact close to the border of the neighboring cluster. It is apparent from this phenomenon, and important to stress here, that the clustering technique is a \emph{tool}. The field distribution data is not categorical, so we expect superposed solutions which are badly represented by \enquote{pure} modes. That said, these intermediate parts are small, as it is seen from the saturation distributions, so that the clustering is still a valid and effective approach.

\subsection*{Silhouette analysis and the number of clusters}

Before we investigate the field distribution prototypes, the quality of the clustering itself is evaluated using a mathematical analysis in the following. This can be done using the silhouette coefficients, using a scheme known as \emph{silhouette analysis}\cite{Rousseeuw1987}. \Cref{fig:silhouettes} depicts so-called \emph{silhouette plots} for each combination using the same column order as in \cref{fig:enhancement_vs_clustering}. In each of the plots, the silhouette coefficients for each sample are plotted as a bar in $x$-direction with a length corresponding to its value (negative values point into the $-x$-direction). The samples are sorted by their silhouette coefficients, with smaller values being located at smaller $y$-positions. In addition, the samples are grouped for each cluster $k$ and color-coded using the same colors as in \cref{fig:enhancement_vs_clustering} (lower row). The red dashed lines mark the average of all silhouette coefficients, which is a measure for the absolute quality of the representation denoted as \emph{silhouette score}. The results are the typical \enquote{sails} or \enquote{shark fins}. The width of each fin in the silhouette plots is proportional to the area of the correspondingly labelled points in the classification maps.

Considering the distribution of the silhouette coefficients, fins which are not too sharp are observed, i.e.~having broad plateaus of high silhouette coefficients. There is only a minimum number of values with negative coefficients. Both arguments together give a validation for the fact that the number of clusters is not underestimated: negative values would occur if there were to few clusters, leaving back samples which do not fit in one of the classes ($s\sim-1$). Too many clusters could be identified by a large fluctuation in the fin widths. But this does not fully apply here, as the areas occupied by the bands and the residual parts are unequal. Therefore, equally broad classes are not expected. A slightly too large number of clusters can be seen as unproblematic, because it would basically subdivide the radiation mode regions further, which are of limited relevance for the interpretation. Another point that suggests a good representation is that there are few clusters with below average silhouette scores. Using this reasoning, the optimum number of clusters was determined for each case by comparing the silhouette plots for different number of clusters (results omitted).

\subsection*{Field distribution prototypes}

\begin{figure*}[t]
	\centering \includegraphics[scale=0.9]{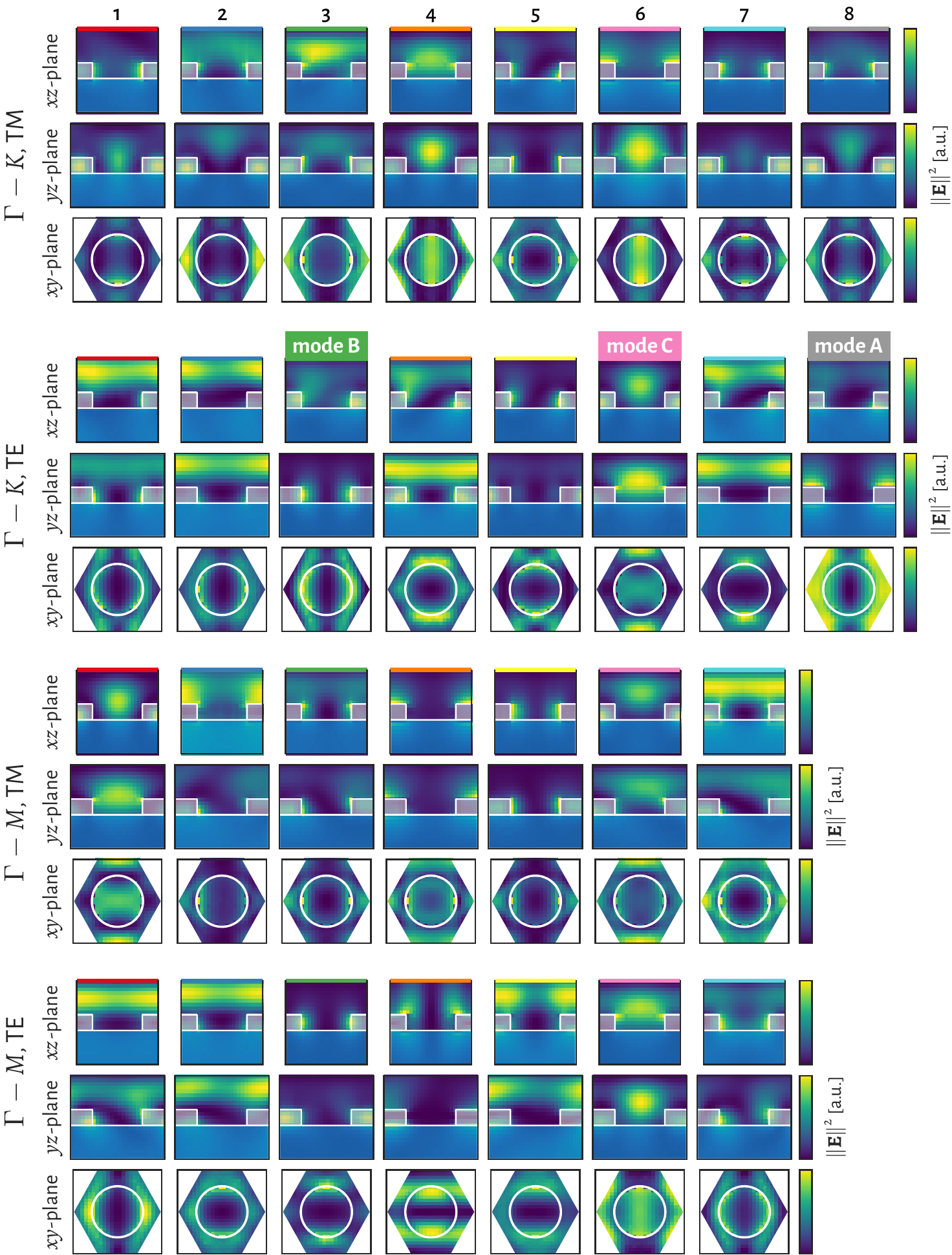}
	\caption{{\setstretch{1.0}\footnotesize\textbf{Prototype maps for the different direction/polarization combinations.} For each direction/polarization combination (4 horizontal panels) a prototype (i.e.~cluster center) map is shown, which consists of 3 \textbf{rows} for the $xz$, $yz$ and $xy$ planes, respectively (from top to bottom). Each of these rows has a number of \textbf{columns} that accounts for the number of clusters, and each column has a colored edge in the top-most row that corresponds to the color used for that label in the classification maps shown in the lower row of \cref{fig:enhancement_vs_clustering}. The cluster label is further given in the title of the $xz$-row. Each distribution plot depicts the electric field energy distribution $\norm{\vec{E}}^2$ in the respective plane. The distribution plots feature semitransparent markings for the glass superstrate (blue) and the silicon of the PhC (gray) in the case of $xz$ and $yz$; and a white circle indicating the hole circumference in the case of $xy$. (Color scales do not give absolute values, as the prototypes are based on normalized data.)}} \label{fig:prototypes}
\end{figure*}

As the second essential outcome, the clustering procedure yields the field distribution prototypes. As the input data for the GMM algorithm has been electric field values on three planes, namely $xy$, $xz$ and $yz$, the prototype data is available on these planes as well. The prototypes for all clusters of each combination, and on all three planes, are depicted in \cref{fig:prototypes}. For each direction/polarization combination a \emph{prototype map} is shown in one of the four horizontal panels. Each panel consists of three rows for the $xz$, $yz$ and $xy$ planes, respectively (from top to bottom). Each of these rows has the same number of columns that accounts for the number of clusters, and each column has a colored edge in the top-most row that corresponds to the color used for that label in the classification maps shown in the lower row of \cref{fig:enhancement_vs_clustering}. The cluster label is further given in the title of the $xz$-row. Each distribution plot depicts the electric field energy distribution $\norm{\vec{E}}^2$ in the respective plane. The distribution plots further feature semitransparent markings for the glass superstrate (blue) and the silicon of the PhC (gray) in the case of $xz$ and $yz$; and a white circle indicating the hole circumference in the case of $xy$. Recall that the color scales do not give absolute values, as the prototypes are based on normalized data and, therefore, cannot be compared with respect to their absolute amplitudes.

For each prototype, the field energy plots on the three planes give a notion of the 3D field energy distribution. The solutions with the same label (color) in the classification maps of \cref{fig:enhancement_vs_clustering} all share this distribution type. Lower saturations quantify how much the individual solutions deviate from the prototype. Clusters that correspond to leaky mode bands with strong field enhancement, such as cluster 6 of the $\Gamma-K$, TM case (pinkish), have strongly localized energy distributions (see panel 1, column 6 in \cref{fig:prototypes}). In contrast, clusters that belong to radiation modes have energy distributions that increase away from the PhC, e.g.~cluster 2 of the $\Gamma-K$, TE case (dark blue, see panel 2, column 2). Further comparisons will be considered below.

% The following fix is from
%     https://tex.stackexchange.com/questions/402627/using-clearpage-after-figure-in-revtex-gives-error-output-routine-didnt-u
% and resolves the error "Output routine didn't use all of \box255."
\makeatletter\onecolumngrid@push\makeatother
\clearpage
\makeatletter\onecolumngrid@pop\makeatother

\subsection*{Putting the pieces together}

To explain the measured fluorescence enhancement effects shown in the upper maps of \cref{fig:enhancement_vs_clustering}, it is necessary to combine all information gained from the numerical analysis. This is, the volume-integrated field energy enhancement maps ($E_+$, \cref{fig:enhancement_vs_clustering}, center row), the classification maps (\cref{fig:enhancement_vs_clustering}, lower row) and the prototype maps (\cref{fig:prototypes}). A \emph{guide} on how the different aspects of the results can be connected to yield a complete interpretation may read as follows:
\begin{enumerate}
	\item Select a feature in the volume-integrated field energy enhancement ($E_+$) maps. For these features the simulation suggests a possible excitation enhancement effect.
	\item Check whether there is an according feature in the experimental fluorescence enhancement ($F_+$) maps.
	\item Afterwards, observe the corresponding region in the classification maps and determine the cluster label from the color using the color bar.
	\item Using this label or color, locate the related column in the prototype map that belongs to the direction/polarization combination (the prototype maps are ordered according to the columns in \cref{fig:enhancement_vs_clustering}, from left to right). Check if the field energy distributions on the three planes can explain the observed fluorescence enhancement (this may necessitate to take into account all cases, because the QD distribution is unknown).
\end{enumerate}
For ease of comprehension, we will analyze the results for selected cases in order of increasing complexity.

\paragraph*{$\Gamma-M$, TM.}

The experimental fluorescence enhancement ($F_+$) map features a single stripe of increased fluorescence with a high contrast. This stripe excellently corresponds to the single leaky-mode band causing a high volume-integrated field enhancement in the $E_+$ map. The classification map reveals this band accordingly with label 4 (orange), for which the field distributions are shown in column 4 of the prototype map. The $xz$ and $yz$ patterns show that the energy of this band is accumulated at the \emph{plateaus} between the holes. As the QDs in this experiment are distributed inside the holes and particularly in an about \SIrange{100}{300}{\nm} thick film on top of the structure, they overlap with the leaky mode volume very well. When observing other columns of the prototype map, there are other interesting patterns which could potentially increase the emission of the QDs, for instance in columns 1 and 5. These two modes gather their energy in the \emph{center of the hole} and at the \emph{flanks}, respectively. However, when looking at the classification maps again, these correspond to the red and yellow regions, which are outside the measurement window. Another important point is given by the patterns of solutions that are related to radiation modes. These would be expected at regions \emph{off} any band, e.g.~in the dark blue and cyan regions with labels 2 and 7. Returning to the prototype maps, these modes in fact have energy distributions that increase with the distance from the PhC surface, but less dominant in case of the the prototype of cluster 2. Moreover, these modes do not fulfill the necessary condition 2 of the guide given above, i.e.~they do not exhibit an integrated field energy enhancement ($E_+$).

\paragraph*{$\Gamma-K$, TM.}

We observe a steep band of high fluorescence and a large field energy enhancement (condition 2) at a similar position. The clustering approach found the band as well, labeled as 6 (pinkish). Column 6 of the prototype map shows that this band localizes its energy at the center of the hole (slightly above the PhC in $z$-direction) but also at the plateaus in the $xz$ plane. It therefore potentially affects QDs relatively independently of whether they are gathered inside the hole or on the plateaus. Again, as the emitters are distributed inside the hole and in a bulk film on top of the structure in the experiment, the overlap with the leaky mode field distribution is good. Returning to the clustering, it is seen that a second, much broader band running from top left to bottom right is seen in the $E_+$ maps. The classification maps reveal that the field distribution of this band undergoes a change when crossing the pinkish band, from label 7 (cyan) to label 1 (red). This band is basically not seen in the fluorescence enhancement.%, \emph{but} there is a very bright region above \ang{50} in the fluorescence. When keeping to the guide given above, it follows from step 2 that this feature might not be explained from the numerical data, because there is no field enhancement. However, if looking closely there is a kind of \enquote{transition boundary} in the $E_+$ map at almost exactly the outline of the bright fluorescence region. In the classification map, there is a feature at this location as well: a kind of \enquote{stroke} across the red, gray and dark blue regions. In fact, the prototype maps show that these clusters have similar energy distributions compared to the pinkish band. Although this still does not explain the fluorescence enhancement, it shows that there is in fact an effect at exactly this position which could potentially explain it. It is conceivable that this mode couples very effectively to the QDs and that the field enhancement in the real system is higher than suggested by the simulation.

\begin{figure*}[t]
	\centering \includegraphics{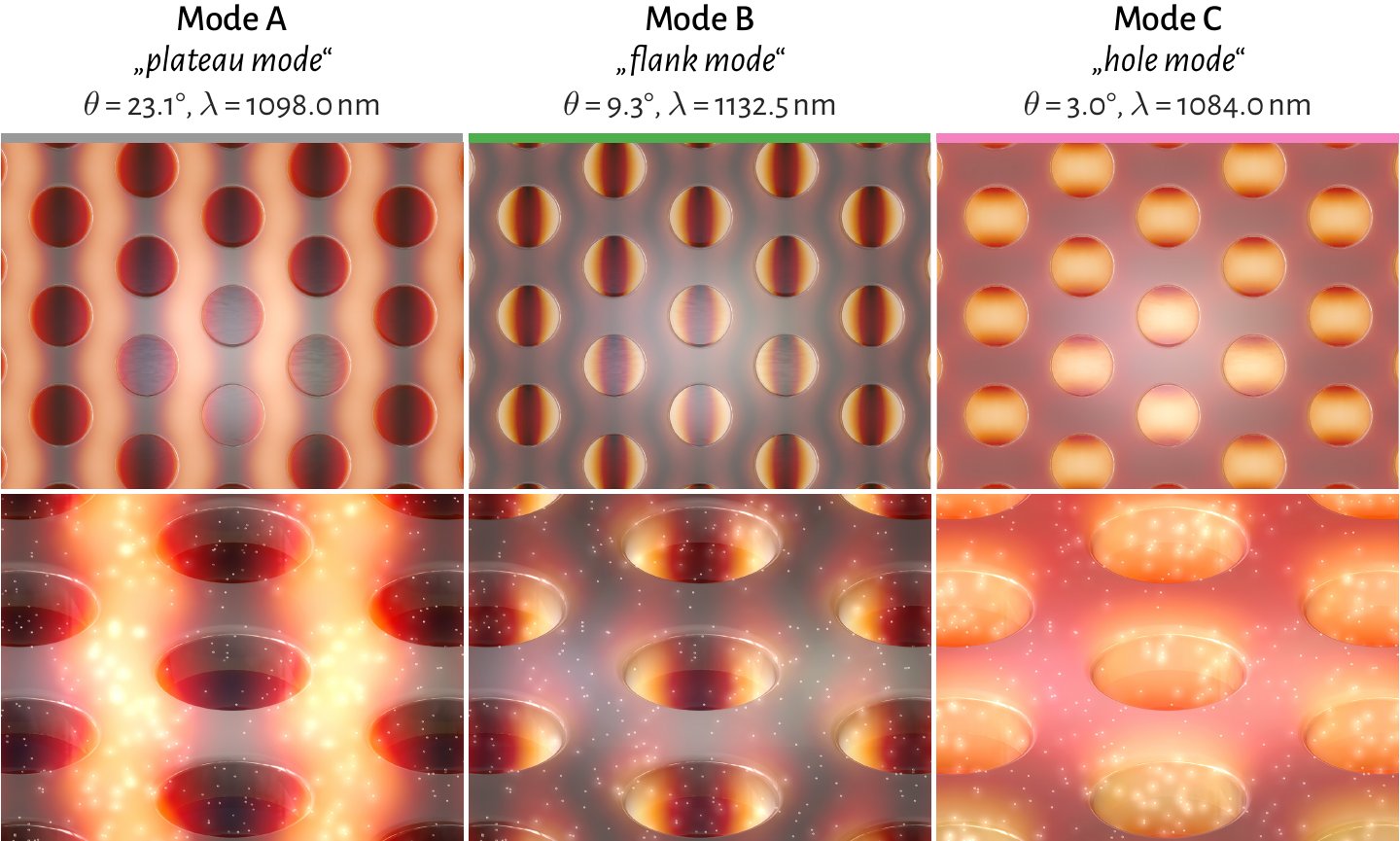}
	\caption{\textbf{Full-3D volume renderings of selected modes for the $\Gamma-K$, TE case.} Semi-artistic ray tracing images depicting multiple periods of the photonic crystal as a grayish material. The \textbf{upper row} shows a topview of the full-3D $E$-field energy density, color-coded using a heat map (not comparable between the figures). The \textbf{lower row} shows a closer view and indicates the same random distribution of quantum dots (bright small spheres), emitting white light with an intensity proportional to the field energy density at their specific positions. The \textbf{columns} relate to three different modes of the $\Gamma-K$/TE case, denoted as modes A, B and C, as marked in \cref{fig:prototypes}. The modes are the actual solutions from the finite-element solver that have the smallest deviations from the assigned prototype (i.e.~cluster center), determined using the silhouette coefficients. Incident angle $\theta$ and wavelength $\lambda$ for each mode are given in the headings. The figures use real physical proportions.} \label{fig:povray_modes}
\end{figure*}

\paragraph*{$\Gamma-K$, TE.}

The TE cases both feature a more complicated band structure. In the $\Gamma-K$, TE case, there are two very clear bands that show anticrossing, a steeper band crossing the complete wavelength range from roughly \ang{20} to \ang{40}, and a shallower one coming from top left. The former is very clearly seen in the clustering by the gray region with label 8. From the prototype map it is observed that this band has a node along the $x$-direction and concentrates its energy at the flanks and the plateaus in $y$-direction. The energy distribution is therefore comparable to the one of the orange band in the $\Gamma-M$, TM case, which is clearly seen in the experimentally measured fluorescence as well. The shallower mentioned band undergoes a transition from the green cluster (label 3) to the red cluster (label 1) in the classification maps. For the green parts the energy is strongly localized at the flanks, while the red one can be identified with a radiation mode. The energy is therefore less well confined to the surface in the red case, which is exactly seen in the fluorescence maps, where only at the location of a broad green region a fluorescence enhancement can be observed, but no enhancement is seen at the prosecution of the mode at higher incident angles.

\paragraph*{$\Gamma-M$, TE.}

The $E_+$-maps cover two bands that show anticrossing at the long wavelength/steep angle end of the data window (top left). The lower band and most parts of the interaction zone are labeled as cluster 3 (green), while the upper band has label 7 (cyan). The green band has a node in the $yz$ plane, while the energy in $x$-direction is strongly localized at the flanks. It is therefore likely to be seen in the fluorescence enhancement, when comparing to the previous results. However, a fluorescence enhancement for the green band is mainly seen in the interaction zone. This can be explained when considering the energy distribution of the cyan band: this band does not have the node in the $yz$ plane. In the interaction zone, the two bands basically overlap, so that the node is partially erased. Therefore, a stronger effect in the fluorescence enhancement is expected, just as it is observed. Interestingly, there is moreover a small effect at $\approx\SI{1080}{\nm}$ for large angles in the $F_+$ map. The clustering reveals another band with strong localization at the flanks: the orange region with label 4 (and the similar yellow one which couples more strongly to the radiation modes). It is likely that the measured effect is actually caused by this band, as it has a similar energy distribution as other bands which are clearly seen.

To give a clear idea of the 3D energy density distribution for three selected modes, and also to show how well the clusters match the actual physical fields, \cref{fig:povray_modes} shows full-3D renderings\cite{POVRay2013}. The images depict multiple periods of the photonic crystal as a grayish metal-like material, without showing a superspace material. The upper row shows a topview of the volume-rendered electric field energy density color-coded using a heat map, which is not comparable between the figures. The lower row shows a closer view and indicates a random distribution of QDs as bright small spheres, emitting white light with an intensity proportional to the field energy density at their specific positions. The QD distribution is the same for all three images. The columns relate to three different modes of the $\Gamma-K$, TE case, denoted as A, B and C. They correspond to clusters 8, 3 and 6, as also marked in \cref{fig:prototypes}. The modes are the actual solutions from the finite-element solver that have the smallest deviations from the assigned prototype (i.e.~cluster center), determined using the silhouette coefficients. Incident angle $\theta$ and wavelength $\lambda$ for each mode are given in the headings. Note that these images have an illustrative character, but can be very helpful to imagine the actual physical situation. Modes A and B are the ones which have been discussed recently, and it is clearly seen that the former concentrates its energy at the \emph{plateaus}, while the latter has high energy densities at the \emph{flanks} of the holes. A third type is shown with mode C, which focusses the energy directly \emph{inside the holes}. The illustrations in the lower row give a notion of how these modes activate different QDs, depending on their position. Only a small density of QDs is used for the images for purposes of visibility, and they are randomly distributed in a layer that fills the holes and extents \SI{100}{\nm} in $z$-direction. Mode A very efficiently excites QDs at the plateaus, just as expected, while modes B and C do the same at the flanks and inside the holes, respectively. Consequently, these renderings completely confirm the results of the clustering approach.

\section*{Discussion}

%Proposal ChB:
The aim of the numerical approach presented here is the systematic identification of suitable leaky modes of nanophotonic structures for interaction with near-surface emitters. For instance, (a) a monolayer of emitting species attached at the surface of the nanophotonic structure is expected to strongly interact with a leaky mode with shallow field distribution. In contrast, an experiment with (b) emitters in a coating on top of the photonic nanostructure, the interaction with a shallow leaky mode will be rather small due to the limited spatial overlap of the mode volume with the emitting material. Here, a leaky mode with an energy density enhancement in a large volume outside the photonic nanostructure would be better suited. In a third scenario, (c), emitters fill the voids of the photonic nanostructure, for example if they are solved in a liquid solution and dropped onto the structure. In that case, leaky modes with strong field enhancement inside these voids are expected to cause the strongest effects. In the chosen dataset, the fluorescence enhancement experiment of PbS quantum dots on a silicon PhC slab in nanohole geometry, the distribution of QDs resembles a mixture of case (b) and (c). The clustering technique revealed that the modes which have the best spatial overlap with the QD distribution effectively cause the strongest fluorescence enhancement effects in the measurements.

%The prior discussion showed a very consistent explanation of nearly all the effects observed in the fluorescence enhancement experiment. Even for the most complicated case of $\,\Gamma-M$, TE a reasonable interpretation of the elongated bright spot (or rather two very close spots) has been found by a superposition of modes; and moreover light was shed on the very small effect in the bottom right corner. Similarly, it was possible to explain why the shallower band in the $\Gamma-K$, TE case does not cause a strong fluorescence beyond the anticrossing point. And even for the bright spot at shallow angles in the $\Gamma-K$, TM case a potential explanation was found.

In the previous study\cite{Barth2017} we used a selection of a small number of points for which the field energy distributions was analyzed. The clustering technique confirmed the results that were achieved this way, but it also helps to explains complicated details, e.g.~as caused by the superposition of two modes. Therefore, the clustering approach gave a much more coherent and detailed explication of the underlying physical phenomena. It emphasizes the interesting parts automatically and systematically, e.g.~by revealing regions of rapidly changing field distributions through individual clusters, or through large deviations from the assigned prototype. Moreover, the clustering technique seems to be applicable to even more complicated cases, e.g.~in windows with more bands for which an analysis using selected points is not reasonable any more.

%\subsection*{Relevance of the presented analysis technique}

The presented technique composed of (i) the field energy enhancement maps and (ii) the 3D electric field distribution clustering provides a versatile tool for the analysis and design of photonic nanostructures for applications that utilize near-field enhancement effects for increased emission. For any known distribution of near-surface emitters that should be affected by leaky modes, optimum values for all relevant parameters can in principle be determined. It is e.g.~possible to define a wavelength range for the excitation of the emitters by considering their absorption properties, and to numerically calculate the field energy enhancement $E_+$ and field values in 3D for clustering (as provided in the center and lower row of \cref{fig:enhancement_vs_clustering}, here). By choosing the mode with the largest spatial overlap of high field energy with the emitter distribution from the prototypes (as in \cref{fig:prototypes}), an optimum mode can systematically be determined.
%\begin{enumerate}
%	\item Define a wavelength range for the excitation of the emitters by considering their absorption properties, and possibly also their emission properties. The latter is relevant if an interaction of the emitted radiation with leaky modes is to be avoided (or even desired, if extraction enhancement effects should be utilized at the same time).
%	\item By using FEM simulations with a model that fits the actual final system, perform a parameter scan over the incident wavelengths, angles and polarizations and calculate the field energy enhancement $E_+$ and field values in 3D for clustering. This process results in the information as given in \cref{fig:enhancement_vs_clustering} (center and lower row) for the present case.
%	\item Choose the mode that causes the largest overlap of high field energy with the known emitter distribution from a spatial analysis of the prototypes (\cref{fig:prototypes}). This step will define the high-symmetry direction and polarization, supposed there is a mode that is a clear optimum.
%	\item Determine the incident wavelength and angle $\theta$ by choosing the configuration with the largest field energy enhancement $E_+$ for the selected mode. 
%\end{enumerate}
This process can moreover be repeated for possible geometrical parameters of the photonic nanostructure, e.g.~the lattice constant, slab thickness or hole radius. Alternatively, if the geometrical parameters should be varied extensively, the technique could be applied for an initial set of geometrical parameters to select a potential mode and to reduce the wavelength and angle window. Successively, only the field energy enhancement $E_+$ may be calculated in the scan over the possible geometrical parameters to determine the absolute maximum of the enhancement.

The clustering technique is extremely flexible. It is not limited to uniformly sampled feature spaces as shown in our example application. It would also have been possible to choose arbitrary snapshot points in the $\theta$-$\lambda$ space, e.g.~with a higher density in regions of high field energy enhancement $E_+$. It is further not limited to the shown number of feature parameters, i.e.~we could have added a variation of the hole diameter or other geometrical parameters as well. But the method is even more powerful, because the trained classifier can be used to classify field distributions that it has not \enquote{seen} yet, known as \emph{prediction}. In contrast to the clustering itself, this is a computationally cheap process, and the classifier can even be persistently stored on disk for later use. To make these considerations more clear, it would have been possible to choose a smaller number of possibly non-uniformly sampled points in the $\theta$-$\lambda$ space for efficient clustering. The silhouette analysis can be used to make sure that the number of samples is sufficient to reach an appropriate clustering result. From this clustering the prototype field distributions can be derived and the classifier can be stored to disk. Afterwards, an e.g.~uniform scan over $\theta$, $\lambda$, and other parameters that are expected to not change the field distributions considerably, (e.g.~hole diameter, slab thickness, refractive indices, \ldots) could be performed. The resulting new solutions could then be assigned to the prototypes using the classifier from disk with minimal computational effort.

Numerous applications could benefit from these optimization abilities. In the field of biosensing, photonic nanostructures have become an important platform for e.g.~label-free biosensing or for the enhancement of the output of photon emitting tags used in the life sciences and \emph{in vitro} diagnostics. A recent review article\cite{Cunningham2016} shows that nanophotonic enhanced biosensors are yet extremely relevant, even commercially and potentially on industrial scale. Exploiting leaky modes with large Q-factors enables for narrow bandwidths ($<\SI{1}{\nm}$) and extremely high sensitivities, e.g.~for detection of disease biomarkers in serum with concentrations of $\sim\SI{1}{\pico\gram\per\ml}$. The numerous applications that are described in the mentioned review article have in common that the nanophotonic structure is designed for a very specific mode, i.e.~a specific illumination condition and a determinable distribution of the molecules/cells/virus particles in question. This is where the technique presented here could be utilized for a \emph{systematic} optimization in the design process, and hence to further increase the sensitivities of related sensors. Photon upconversion\cite{Schulze2014a,Park2015} in biomedical imaging and solar energy is another application that could benefit from the discussed all-numerical design abilities.  Recent publications\cite{Wu2015b,Wu2015c} demonstrate upconversion using thin emitter layers, which as well could potentially be improved using specifically tailored nanophotonic structures.

In summary, we have developed a numerical method that allows to systematically optimize nanophotonic structures pertaining to the 3D field distribution and field energy enhancement of modes. The method uses a combination of FEM simulations and post-processing using machine learning clustering. We showcased the modelling power of the method by explaining experimentally measured fluorescence enhancement of QDs on a photonic crystal slabs surface. The method yielded information that was not easily accessible using e.g.~a visualization-based analysis for selected parameter combinations, and which allowed to fully explain the experimental results. Consequently, the presented technique could be of great avail for applications that utilize effects that depend on the spatial field distribution of nanophotonic modes, such as in the fields of biosensing\cite{Park2015,Cunningham2016}, quantum dot solar cells\cite{Carey2015a,Kamat2008a,AkihiroKojimaKenjiroTeshimaYasuoShirai2009,Kundu2017}, or up-conversion in solar energy\cite{Wu2015b,Wu2015c,Schulze2014a}.

% =============================================================================
\section*{Methods}\label{sec:methods}

\subsection*{Clustering of electric field data}

The clustering is executed on an input matrix $X$ of shape $N_s\times{}N_f$, where $N_s$ is the number of \emph{samples} and $N_f\,$ the number of \emph{features}. A sample is the solution for a specific set of input parameters, in our case incident angle $\theta$ and wavelength $\lambda$. The features, in the present case, are absolute values of the electric field components $E_{j}$ with $j\in\{x,y,z\}$ for a number of points $\vec{r}_i\in\mathbb{R}^3$, i.e.~of the form $\abs{E_{j}(\vec{r}_i)}$. Consequently, if the field is evaluated at $N_p$ points, these are $N_f=3N_p$ features. To avoid exporting the electric field on a full Cartesian grid in 3D, which would cause huge amounts of data when trying to achieve a reasonable resolution, data is only exported on the symmetry planes marked in \cref{fig:overview}(c), respectively. More symmetry planes could be used as well, but based on these three planes a reasonable classification can be reached, as tested using smaller data sets and comparing to a full 3D field output. A field pattern of a single simulation holds data for each of the 3 spatial directions, and for each component $j$ of the electric field (altogether a 4D data set). As each sample $X_i$ must be a 1D row vector with observations of single scalar values $x_0,\ldots,x_{N_f\,-1}$, it is necessary to flatten these data sets in always the same way, yielding \enquote{1D representations} of the fields. The data is moreover normalized by scaling each sample to unit norm individually. The field export is performed for each point in each map of \cref{fig:enhancement_vs_clustering}, center row, so that the samples are unique simulations for a given direction/polarization combination, wavelength $\lambda$ and incident angle $\theta$. The number of samples for a single map is given by $N_s=N_\lambda\cdot N_\theta$. To give an expression for the complete input matrix $X$ we abbreviate $\widehat{E}_{j}^{\,i,m,l}=\abs{E_{j}(\vec{r}_i, \theta_m, \lambda_l)}$, where the additional indices $m=0\ldots N_\theta$ and $l=0\ldots N_\lambda$ have been introduced, and where the hat denotes the absolute value and normalization. The input matrix then reads
\begin{widetext}
	\begin{equation}
	X=\begin{pmatrix}
	\widehat{E}_{x}^{\,0,0,0} & \cdots & \widehat{E}_{x}^{\,N_p,0,0} & \cdots & \widehat{E}_{z}^{\,0,0,0} & \cdots & \widehat{E}_{z}^{\,N_p,0,0} \\
	&  &  & \vdots &  &  &  \\[\medskipamount]
	
	\widehat{E}_{x}^{\,0,N_\theta,0} & \cdots & \widehat{E}_{x}^{\,N_p,N_\theta,0} & \cdots & \widehat{E}_{z}^{\,0,N_\theta,0} & \cdots & \widehat{E}_{z}^{\,N_p,N_\theta,0}\\[\medskipamount]
	&  &  & \vdots &  &  &  \\[\medskipamount]
	
	\widehat{E}_{x}^{\,0,N_\theta,N_\lambda} & \cdots & \widehat{E}_{x}^{\,N_p,N_\theta,N_\lambda} & \cdots & \widehat{E}_{z}^{\,0,N_\theta,N_\lambda} & \cdots & \widehat{E}_{z}^{\,N_p,N_\theta,N_\lambda}
	\end{pmatrix}
	\end{equation}
\end{widetext}
For the wavelength and angle resolution values of \SI{0.5}{\nm} and \ang{0.3} have been used, respectively. For each clustering procedure the input matrix $X$ had a size of $N_s\times{}N_f=\num{47034}\times\num{8616}$. This is a comparably large problem size, especially because the large feature dimensionality ($N_f$), so that the procedure took more than 10 hours on a hexa-core workstation with roughly 40\,GB of memory consumption.

\subsection*{Gaussian mixture model clustering}

Simple clustering techniques, such as the $k$-means algorithm\cite{Bishop2013}, can be extremely robust, but also have their disadvantages. E.g.~$k$-means assumes that the clusters are circular, i.e.~representable by a (hyper-)sphere in feature space. The center of this sphere defines the cluster center (i.e.~prototype), while the radius acts as a \emph{hard} boundary used to decide which samples belong to the cluster. In contrast, the GMM\cite{Bishop2013,Mining2009} is a so-called \emph{soft} method. That is, a \enquote{score} for each cluster is assigned to the samples, which account for the probability that the sample belongs to a specific cluster. In GMM clustering, the clusters are represented by Gaussian distributions of the dimensionality of the features space (i.e.~$N_f$).

In general, a superposition of $N$ multivariate Gaussian distributions of the form
\begin{equation}
p(\vec x) = \sum_{i=1}^N c_i \mathcal{N}_i(\vec x)
\label{eq:gaussian_mixture}
\end{equation}
can be used to approximate almost any continuous density to arbitrary accuracy (this is intuitive with 1D Gaussians, which can fit almost any 1D signal if enough Gaussians are superimposed). Here, the $\mathcal{N}_i(\vec{x})$ are multivariate Gaussian distributions of the form\cite{Bishop2013}
\begin{equation}
\mathcal{N}(\vec{x}) = \frac{1}{\mathstrut{(2\pi)^{D/2}|\mat{\Sigma}|}^{1/2}}\exp\left(-\frac{\left(\vec{x}-\vec{\mu}\right)^T}{2}\mat{\Sigma}^{-1}\left(\vec{x}-\vec{\mu}\right)\right)
\label{eq:gaussian_multi}
\end{equation}
for a $D$-dimensional vector $\vec{x}$, the $D$-dimensional mean-vector $\vec{\mu}$, and the $D\times D$ covariance matrix $\mat{\Sigma}$ with determinant $|\mat{\Sigma}|$. \Cref{eq:gaussian_mixture} is called a Gaussian mixture, the $\mathcal{N}_i(\vec x)$ are called components of the mixture, and the $c_i$ are weight factors. Loosely speaking, the distribution of sample points is \enquote{fitted} using a set of high-dimensional Gaussians. A GMM can therefore represent much more complex data sets and can be seen as a generalization of the $k$-means algorithm for non-circular clusters. One can imagine that it would be straightforward to fit the multivariate Gaussians to a data set for which the labels are known. With unlabeled data the case is more difficult, and enforces to take into account another step. In the literature, this problem is commonly denoted as to find out which (latent) component is \enquote{responsible} for a certain sample, -- which is somehow a different way of asking to which cluster the sample belongs. But it underlines that the GMM clustering is a probabilistic approach, because it calculates the probability that the sample was generated by cluster $i$ for \emph{all} clusters. These probabilities, which are also called responsibilities, are simply the weight factors $c_i$ of \cref{eq:gaussian_mixture}. In the implementation that was utilized here, the cluster assignment is solved using a method known as expectation-maximization \cite{Dempster1977,McLachlan1997}. This algorithm starts with a random Gaussian mixture (i.e.~random components), which is typically initialized using a prior application of $k$-means to improve the convergence. In the next step it determines for each sample the probability of being generated by each component of the mixture. Based on these probabilities, the parameters of the Gaussian distributions are fitted to give the best approximation of the data by maximizing their likelihood \cite{Bishop2013}. This process is executed iteratively and is guaranteed to converge to a local optimum.

\subsection*{Solution quality rating using silhouette coefficients}

To give a definition of the silhouette coefficient, let $X_i^k$ be a sample that was assigned to the cluster $k$ and $a(i)$ be the average dissimilarity of $X_i^k$ to all other members $X_{j\neq i}^k$ of this cluster. The measure for the dissimilarity is usually the Euclidian distance. Let $d(i,m)$ be the average dissimilarity of $X_i^k$ to all members of the cluster $m\neq k$ and $b(i)$ be the minimum of $d(i,m)$ for these clusters, i.e.
$$b(i)=\min_{m\neq k}d(i,m).$$
The cluster $m$ for which this minimum is obtained is called the neighboring cluster of $X_i$. If the number of clusters $N_k$ is $>1$, we can define the silhouette coefficient $s(i)$ for the sample $X_i$ by
\begin{equation}
\begin{split}
s(i) &= \frac{b(i) - a(i)}{\max\{a(i),b(i)\}}\\[.5\baselineskip]
     &= \begin{cases}
1-a(i)/b(i), & \mbox{if\;\;} a(i) < b(i) \\
0,  & \mbox{if\;\;} a(i) = b(i) \\
b(i)/a(i)-1, & \mbox{if\;\;} a(i) > b(i)
\end{cases}.
\label{eq:ML_silhouette}
\end{split}
\end{equation}
From this definition it is seen that the silhouette coefficient $s$ is in the range $-1\leq s\leq 1$. Values near \num{+1} indicate that the sample is far away from the neighboring cluster and accordingly fits well into its own cluster. A value of \num{0} indicates that the sample is on or very close to the boundary between its own and the neighboring cluster, and negative values indicate that it might have been assigned to the wrong cluster. A sorted diagram of all silhouette coefficients can thus be used to visualize the representation quality of a clustering. In addition, the average silhouette coefficient for all samples -- usually denoted as \emph{silhouette score} -- can be used to compare the representation quality for different clusterings, e.g.~using different $N_k\,$-values. It hence even provides a single numeric value for solution quality assessment.

%Explanations on how to use these diagrams will be provided in the corresponding sections, for instance in \cref{subsec:opex_clustering} (\cref{fig:opex_silhouettes}).

%%%%%%%%%%%%%%%%%%%%%%%%%%%%%%%%%%%%%%%%%%%%%%%%%%%%%%%%%%%%%%%%%%%%%
\begin{acknowledgments}

The authors thank Klaus J{\"{a}}ger from Helmholtz-Zentrum Berlin for useful discussions. The German Federal Ministry of Education and Research is acknowledged for funding the research activities of the Nano-SIPPE group within the program NanoMatFutur (No. 03X5520). Further we acknowledge support by the Einstein Foundation Berlin through ECMath within subproject SE6. Parts of the results were obtained at the Berlin Joint Lab for Optical Simulations for Energy Research (BerOSE) of Helmholtz-Zentrum Berlin f\"{u}r Materialien und Energie, Zuse Institute Berlin and Freie Universit\"{a}t Berlin.

\end{acknowledgments}

%%%%%%%%%%%%%%%%%%%%%%%%%%%%%%%%%%%%%%%%%%%%%%%%%%%%%%%%%%%%%%%%%%%%%
%merlin.mbs aipnum4-1.bst 2010-07-25 4.21a (PWD, AO, DPC) hacked
%Control: key (0)
%Control: author (8) initials jnrlst
%Control: editor formatted (1) identically to author
%Control: production of article title (0) allowed
%Control: page (1) range
%Control: year (1) truncated
%Control: production of eprint (0) enabled
%


\begin{thebibliography}{70}%
	\makeatletter
	\providecommand \@ifxundefined [1]{%
		\@ifx{#1\undefined}
	}%
	\providecommand \@ifnum [1]{%
		\ifnum #1\expandafter \@firstoftwo
		\else \expandafter \@secondoftwo
		\fi
	}%
	\providecommand \@ifx [1]{%
		\ifx #1\expandafter \@firstoftwo
		\else \expandafter \@secondoftwo
		\fi
	}%
	\providecommand \natexlab [1]{#1}%
	\providecommand \enquote  [1]{``#1''}%
	\providecommand \bibnamefont  [1]{#1}%
	\providecommand \bibfnamefont [1]{#1}%
	\providecommand \citenamefont [1]{#1}%
	\providecommand \href@noop [0]{\@secondoftwo}%
	\providecommand \href [0]{\begingroup \@sanitize@url \@href}%
	\providecommand \@href[1]{\@@startlink{#1}\@@href}%
	\providecommand \@@href[1]{\endgroup#1\@@endlink}%
	\providecommand \@sanitize@url [0]{\catcode `\\12\catcode `\$12\catcode
		`\&12\catcode `\#12\catcode `\^12\catcode `\_12\catcode `\%12\relax}%
	\providecommand \@@startlink[1]{}%
	\providecommand \@@endlink[0]{}%
	\providecommand \url  [0]{\begingroup\@sanitize@url \@url }%
	\providecommand \@url [1]{\endgroup\@href {#1}{\urlprefix }}%
	\providecommand \urlprefix  [0]{URL }%
	\providecommand \Eprint [0]{\href }%
	\providecommand \doibase [0]{http://dx.doi.org/}%
	\providecommand \selectlanguage [0]{\@gobble}%
	\providecommand \bibinfo  [0]{\@secondoftwo}%
	\providecommand \bibfield  [0]{\@secondoftwo}%
	\providecommand \translation [1]{[#1]}%
	\providecommand \BibitemOpen [0]{}%
	\providecommand \bibitemStop [0]{}%
	\providecommand \bibitemNoStop [0]{.\EOS\space}%
	\providecommand \EOS [0]{\spacefactor3000\relax}%
	\providecommand \BibitemShut  [1]{\csname bibitem#1\endcsname}%
	\let\auto@bib@innerbib\@empty
	%</preamble>
	\bibitem [{\citenamefont {Jordan}\ and\ \citenamefont
		{Mitchell}(2015)}]{Jordan2015}%
	\BibitemOpen
	\bibfield  {author} {\bibinfo {author} {\bibfnamefont {M.~I.}\ \bibnamefont
			{Jordan}}\ and\ \bibinfo {author} {\bibfnamefont {T.~M.}\ \bibnamefont
			{Mitchell}},\ }\bibfield  {title} {\enquote {\bibinfo {title} {{Machine
					learning: Trends, perspectives, and prospects}},}\ }\href {\doibase
		10.1126/science.aaa8415} {\bibfield  {journal} {\bibinfo  {journal}
			{Science}\ }\textbf {\bibinfo {volume} {349}},\ \bibinfo {pages} {255--260}
		(\bibinfo {year} {2015})}\BibitemShut {NoStop}%
	\bibitem [{\citenamefont {Just}\ \emph {et~al.}(2017)\citenamefont {Just},
		\citenamefont {Pan}, \citenamefont {Cherkassky}, \citenamefont {McMakin},
		\citenamefont {Cha}, \citenamefont {Nock},\ and\ \citenamefont
		{Brent}}]{Just2017}%
	\BibitemOpen
	\bibfield  {author} {\bibinfo {author} {\bibfnamefont {M.~A.}\ \bibnamefont
			{Just}}, \bibinfo {author} {\bibfnamefont {L.}~\bibnamefont {Pan}}, \bibinfo
		{author} {\bibfnamefont {V.~L.}\ \bibnamefont {Cherkassky}}, \bibinfo
		{author} {\bibfnamefont {D.~L.}\ \bibnamefont {McMakin}}, \bibinfo {author}
		{\bibfnamefont {C.}~\bibnamefont {Cha}}, \bibinfo {author} {\bibfnamefont
			{M.~K.}\ \bibnamefont {Nock}}, \ and\ \bibinfo {author} {\bibfnamefont
			{D.}~\bibnamefont {Brent}},\ }\bibfield  {title} {\enquote {\bibinfo {title}
			{{Machine learning of neural representations of suicide and emotion concepts
					identifies suicidal youth}},}\ }\href {\doibase 10.1038/s41562-017-0234-y}
	{\bibfield  {journal} {\bibinfo  {journal} {Nature Human Behaviour}\ }\textbf
		{\bibinfo {volume} {1}},\ \bibinfo {pages} {911--919} (\bibinfo {year}
		{2017})}\BibitemShut {NoStop}%
	\bibitem [{\citenamefont {Gun{\v{c}}ar}\ \emph {et~al.}(2018)\citenamefont
		{Gun{\v{c}}ar}, \citenamefont {Kukar}, \citenamefont {Notar}, \citenamefont
		{Brvar}, \citenamefont {{\v{C}}ernel{\v{c}}}, \citenamefont {Notar},\ and\
		\citenamefont {Notar}}]{Guncar2018}%
	\BibitemOpen
	\bibfield  {author} {\bibinfo {author} {\bibfnamefont {G.}~\bibnamefont
			{Gun{\v{c}}ar}}, \bibinfo {author} {\bibfnamefont {M.}~\bibnamefont {Kukar}},
		\bibinfo {author} {\bibfnamefont {M.}~\bibnamefont {Notar}}, \bibinfo
		{author} {\bibfnamefont {M.}~\bibnamefont {Brvar}}, \bibinfo {author}
		{\bibfnamefont {P.}~\bibnamefont {{\v{C}}ernel{\v{c}}}}, \bibinfo {author}
		{\bibfnamefont {M.}~\bibnamefont {Notar}}, \ and\ \bibinfo {author}
		{\bibfnamefont {M.}~\bibnamefont {Notar}},\ }\bibfield  {title} {\enquote
		{\bibinfo {title} {{An application of machine learning to haematological
					diagnosis}},}\ }\href {\doibase 10.1038/s41598-017-18564-8} {\bibfield
		{journal} {\bibinfo  {journal} {Scientific Reports}\ }\textbf {\bibinfo
			{volume} {8}},\ \bibinfo {pages} {411} (\bibinfo {year} {2018})}\BibitemShut
	{NoStop}%
	\bibitem [{\citenamefont {Steinegger}\ and\ \citenamefont
		{Soding}(2018)}]{Steinegger2018}%
	\BibitemOpen
	\bibfield  {author} {\bibinfo {author} {\bibfnamefont {M.}~\bibnamefont
			{Steinegger}}\ and\ \bibinfo {author} {\bibfnamefont {J.}~\bibnamefont
			{Soding}},\ }\bibfield  {title} {\enquote {\bibinfo {title} {{Clustering huge
					protein sequence sets in linear time}},}\ }\href {\doibase 10.1101/104034}
	{\bibfield  {journal} {\bibinfo  {journal} {bioRxiv}\ }\textbf {\bibinfo
			{volume} {1}},\ \bibinfo {pages} {104034} (\bibinfo {year}
		{2018})}\BibitemShut {NoStop}%
	\bibitem [{\citenamefont {Chen}\ \emph {et~al.}(2018)\citenamefont {Chen},
		\citenamefont {Juan}, \citenamefont {Tsai},\ and\ \citenamefont
		{Lu}}]{Chen2018}%
	\BibitemOpen
	\bibfield  {author} {\bibinfo {author} {\bibfnamefont {C.-C.}\ \bibnamefont
			{Chen}}, \bibinfo {author} {\bibfnamefont {H.-H.}\ \bibnamefont {Juan}},
		\bibinfo {author} {\bibfnamefont {M.-Y.}\ \bibnamefont {Tsai}}, \ and\
		\bibinfo {author} {\bibfnamefont {H.~H.-S.}\ \bibnamefont {Lu}},\ }\bibfield
	{title} {\enquote {\bibinfo {title} {{Unsupervised Learning and Pattern
					Recognition of Biological Data Structures with Density Functional Theory and
					Machine Learning}},}\ }\href {\doibase 10.1038/s41598-017-18931-5} {\bibfield
		{journal} {\bibinfo  {journal} {Scientific Reports}\ }\textbf {\bibinfo
			{volume} {8}},\ \bibinfo {pages} {557} (\bibinfo {year} {2018})}\BibitemShut
	{NoStop}%
	\bibitem [{\citenamefont {Kan}(2017)}]{Kan2017}%
	\BibitemOpen
	\bibfield  {author} {\bibinfo {author} {\bibfnamefont {A.}~\bibnamefont
			{Kan}},\ }\bibfield  {title} {\enquote {\bibinfo {title} {{Machine learning
					applications in cell image analysis}},}\ }\href {\doibase
		10.1038/icb.2017.16} {\bibfield  {journal} {\bibinfo  {journal} {Immunology
				and Cell Biology}\ }\textbf {\bibinfo {volume} {95}},\ \bibinfo {pages}
		{525--530} (\bibinfo {year} {2017})}\BibitemShut {NoStop}%
	\bibitem [{\citenamefont {Exbrayat}, \citenamefont {Liu},\ and\ \citenamefont
		{Williams}(2017)}]{Exbrayat2017}%
	\BibitemOpen
	\bibfield  {author} {\bibinfo {author} {\bibfnamefont {J.~F.}\ \bibnamefont
			{Exbrayat}}, \bibinfo {author} {\bibfnamefont {Y.~Y.}\ \bibnamefont {Liu}}, \
		and\ \bibinfo {author} {\bibfnamefont {M.}~\bibnamefont {Williams}},\
	}\bibfield  {title} {\enquote {\bibinfo {title} {{Impact of deforestation and
				climate on the Amazon Basin's above-ground biomass during}},}\ }\href
{\doibase 10.1038/s41598-017-15788-6} {\bibfield  {journal} {\bibinfo
		{journal} {Scientific Reports}\ }\textbf {\bibinfo {volume} {7}},\ \bibinfo
	{pages} {1--7} (\bibinfo {year} {2017})}\BibitemShut {NoStop}%
\bibitem [{\citenamefont {Sumpter}\ \emph {et~al.}(2015)\citenamefont
	{Sumpter}, \citenamefont {Vasudevan}, \citenamefont {Potok},\ and\
	\citenamefont {Kalinin}}]{Sumpter2015}%
\BibitemOpen
\bibfield  {author} {\bibinfo {author} {\bibfnamefont {B.~G.}\ \bibnamefont
		{Sumpter}}, \bibinfo {author} {\bibfnamefont {R.~K.}\ \bibnamefont
		{Vasudevan}}, \bibinfo {author} {\bibfnamefont {T.}~\bibnamefont {Potok}}, \
	and\ \bibinfo {author} {\bibfnamefont {S.~V.}\ \bibnamefont {Kalinin}},\
}\bibfield  {title} {\enquote {\bibinfo {title} {{A bridge for accelerating
			materials by design}},}\ }\href {\doibase 10.1038/npjcompumats.2015.8}
{\bibfield  {journal} {\bibinfo  {journal} {npj Computational Materials}\
	}\textbf {\bibinfo {volume} {1}},\ \bibinfo {pages} {15008} (\bibinfo {year}
	{2015})}\BibitemShut {NoStop}%
\bibitem [{\citenamefont {Ramprasad}\ \emph {et~al.}(2017)\citenamefont
	{Ramprasad}, \citenamefont {Batra}, \citenamefont {Pilania}, \citenamefont
	{Mannodi-Kanakkithodi},\ and\ \citenamefont {Kim}}]{Ramprasad2017}%
\BibitemOpen
\bibfield  {author} {\bibinfo {author} {\bibfnamefont {R.}~\bibnamefont
		{Ramprasad}}, \bibinfo {author} {\bibfnamefont {R.}~\bibnamefont {Batra}},
	\bibinfo {author} {\bibfnamefont {G.}~\bibnamefont {Pilania}}, \bibinfo
	{author} {\bibfnamefont {A.}~\bibnamefont {Mannodi-Kanakkithodi}}, \ and\
	\bibinfo {author} {\bibfnamefont {C.}~\bibnamefont {Kim}},\ }\bibfield
{title} {\enquote {\bibinfo {title} {{Machine learning in materials
				informatics: recent applications and prospects}},}\ }\href {\doibase
	10.1038/s41524-017-0056-5} {\bibfield  {journal} {\bibinfo  {journal} {npj
			Computational Materials}\ }\textbf {\bibinfo {volume} {3}},\ \bibinfo {pages}
	{54} (\bibinfo {year} {2017})},\ \Eprint {http://arxiv.org/abs/1707.07294}
{arXiv:1707.07294} \BibitemShut {NoStop}%
\bibitem [{\citenamefont {Russakovsky}\ \emph {et~al.}(2015)\citenamefont
	{Russakovsky}, \citenamefont {Deng}, \citenamefont {Su}, \citenamefont
	{Krause}, \citenamefont {Satheesh}, \citenamefont {Ma}, \citenamefont
	{Huang}, \citenamefont {Karpathy}, \citenamefont {Khosla}, \citenamefont
	{Bernstein}, \citenamefont {Berg},\ and\ \citenamefont
	{Fei-Fei}}]{Russakovsky2015}%
\BibitemOpen
\bibfield  {author} {\bibinfo {author} {\bibfnamefont {O.}~\bibnamefont
		{Russakovsky}}, \bibinfo {author} {\bibfnamefont {J.}~\bibnamefont {Deng}},
	\bibinfo {author} {\bibfnamefont {H.}~\bibnamefont {Su}}, \bibinfo {author}
	{\bibfnamefont {J.}~\bibnamefont {Krause}}, \bibinfo {author} {\bibfnamefont
		{S.}~\bibnamefont {Satheesh}}, \bibinfo {author} {\bibfnamefont
		{S.}~\bibnamefont {Ma}}, \bibinfo {author} {\bibfnamefont {Z.}~\bibnamefont
		{Huang}}, \bibinfo {author} {\bibfnamefont {A.}~\bibnamefont {Karpathy}},
	\bibinfo {author} {\bibfnamefont {A.}~\bibnamefont {Khosla}}, \bibinfo
	{author} {\bibfnamefont {M.}~\bibnamefont {Bernstein}}, \bibinfo {author}
	{\bibfnamefont {A.~C.}\ \bibnamefont {Berg}}, \ and\ \bibinfo {author}
	{\bibfnamefont {L.}~\bibnamefont {Fei-Fei}},\ }\bibfield  {title} {\enquote
	{\bibinfo {title} {{ImageNet Large Scale Visual Recognition Challenge}},}\
}\href {\doibase 10.1007/s11263-015-0816-y} {\bibfield  {journal} {\bibinfo
	{journal} {International Journal of Computer Vision}\ }\textbf {\bibinfo
	{volume} {115}},\ \bibinfo {pages} {211--252} (\bibinfo {year} {2015})},\
\Eprint {http://arxiv.org/abs/1409.0575} {arXiv:1409.0575} \BibitemShut
{NoStop}%
\bibitem [{\citenamefont {Hinton}\ \emph {et~al.}(2012)\citenamefont {Hinton},
	\citenamefont {Deng}, \citenamefont {Yu}, \citenamefont {Dahl}, \citenamefont
	{Mohamed}, \citenamefont {Jaitly}, \citenamefont {Senior}, \citenamefont
	{Vanhoucke}, \citenamefont {Nguyen}, \citenamefont {Sainath},\ and\
	\citenamefont {Kingsbury}}]{Hinton2012}%
\BibitemOpen
\bibfield  {author} {\bibinfo {author} {\bibfnamefont {G.}~\bibnamefont
		{Hinton}}, \bibinfo {author} {\bibfnamefont {L.}~\bibnamefont {Deng}},
	\bibinfo {author} {\bibfnamefont {D.}~\bibnamefont {Yu}}, \bibinfo {author}
	{\bibfnamefont {G.}~\bibnamefont {Dahl}}, \bibinfo {author} {\bibfnamefont
		{A.-r.}\ \bibnamefont {Mohamed}}, \bibinfo {author} {\bibfnamefont
		{N.}~\bibnamefont {Jaitly}}, \bibinfo {author} {\bibfnamefont
		{A.}~\bibnamefont {Senior}}, \bibinfo {author} {\bibfnamefont
		{V.}~\bibnamefont {Vanhoucke}}, \bibinfo {author} {\bibfnamefont
		{P.}~\bibnamefont {Nguyen}}, \bibinfo {author} {\bibfnamefont
		{T.}~\bibnamefont {Sainath}}, \ and\ \bibinfo {author} {\bibfnamefont
		{B.}~\bibnamefont {Kingsbury}},\ }\bibfield  {title} {\enquote {\bibinfo
		{title} {{Deep Neural Networks for Acoustic Modeling in Speech Recognition:
				The Shared Views of Four Research Groups}},}\ }\href {\doibase
	10.1109/MSP.2012.2205597} {\bibfield  {journal} {\bibinfo  {journal} {IEEE
			Signal Processing Magazine}\ }\textbf {\bibinfo {volume} {29}},\ \bibinfo
	{pages} {82--97} (\bibinfo {year} {2012})}\BibitemShut {NoStop}%
\bibitem [{\citenamefont {LeCun}, \citenamefont {Bengio},\ and\ \citenamefont
	{Hinton}(2015)}]{Lecun2015}%
\BibitemOpen
\bibfield  {author} {\bibinfo {author} {\bibfnamefont {Y.}~\bibnamefont
		{LeCun}}, \bibinfo {author} {\bibfnamefont {Y.}~\bibnamefont {Bengio}}, \
	and\ \bibinfo {author} {\bibfnamefont {G.}~\bibnamefont {Hinton}},\
}\bibfield  {title} {\enquote {\bibinfo {title} {{Deep learning}},}\ }\href
{\doibase 10.1038/nature14539} {\bibfield  {journal} {\bibinfo  {journal}
		{Nature}\ }\textbf {\bibinfo {volume} {521}},\ \bibinfo {pages} {436--444}
	(\bibinfo {year} {2015})},\ \Eprint {http://arxiv.org/abs/arXiv:1312.6184v5}
{arXiv:arXiv:1312.6184v5} \BibitemShut {NoStop}%
\bibitem [{\citenamefont {Jain}, \citenamefont {Murty},\ and\ \citenamefont
	{Flynn}(1999)}]{Jain1999}%
\BibitemOpen
\bibfield  {author} {\bibinfo {author} {\bibfnamefont {A.~K.}\ \bibnamefont
		{Jain}}, \bibinfo {author} {\bibfnamefont {M.~N.}\ \bibnamefont {Murty}}, \
	and\ \bibinfo {author} {\bibfnamefont {P.~J.}\ \bibnamefont {Flynn}},\
}\bibfield  {title} {\enquote {\bibinfo {title} {{Data clustering: a
			review}},}\ }\href {\doibase 10.1145/331499.331504} {\bibfield  {journal}
{\bibinfo  {journal} {ACM Computing Surveys}\ }\textbf {\bibinfo {volume}
	{31}},\ \bibinfo {pages} {264--323} (\bibinfo {year} {1999})}\BibitemShut
{NoStop}%
\bibitem [{\citenamefont {Xu}\ and\ \citenamefont {WunschII}(2005)}]{Xu2005}%
\BibitemOpen
\bibfield  {author} {\bibinfo {author} {\bibfnamefont {R.}~\bibnamefont
		{Xu}}\ and\ \bibinfo {author} {\bibfnamefont {D.}~\bibnamefont {WunschII}},\
}\bibfield  {title} {\enquote {\bibinfo {title} {{Survey of Clustering
			Algorithms}},}\ }\href {\doibase 10.1109/TNN.2005.845141} {\bibfield
{journal} {\bibinfo  {journal} {IEEE Transactions on Neural Networks}\
}\textbf {\bibinfo {volume} {16}},\ \bibinfo {pages} {645--678} (\bibinfo
{year} {2005})},\ \Eprint {http://arxiv.org/abs/0912.2303} {arXiv:0912.2303}
\BibitemShut {NoStop}%
\bibitem [{\citenamefont {Aghabozorgi}, \citenamefont {{Seyed Shirkhorshidi}},\
	and\ \citenamefont {{Ying Wah}}(2015)}]{Aghabozorgi2015}%
\BibitemOpen
\bibfield  {author} {\bibinfo {author} {\bibfnamefont {S.}~\bibnamefont
		{Aghabozorgi}}, \bibinfo {author} {\bibfnamefont {A.}~\bibnamefont {{Seyed
				Shirkhorshidi}}}, \ and\ \bibinfo {author} {\bibfnamefont {T.}~\bibnamefont
		{{Ying Wah}}},\ }\bibfield  {title} {\enquote {\bibinfo {title} {{Time-series
				clustering – A decade review}},}\ }\href {\doibase
	10.1016/j.is.2015.04.007} {\bibfield  {journal} {\bibinfo  {journal}
		{Information Systems}\ }\textbf {\bibinfo {volume} {53}},\ \bibinfo {pages}
	{16--38} (\bibinfo {year} {2015})},\ \Eprint {http://arxiv.org/abs/1107.3326}
{arXiv:1107.3326} \BibitemShut {NoStop}%
\bibitem [{\citenamefont {Bhuyan}, \citenamefont {Bhattacharyya},\ and\
	\citenamefont {Kalita}(2014)}]{Garcia-Teodoro2009}%
\BibitemOpen
\bibfield  {author} {\bibinfo {author} {\bibfnamefont {M.~H.}\ \bibnamefont
		{Bhuyan}}, \bibinfo {author} {\bibfnamefont {D.~K.}\ \bibnamefont
		{Bhattacharyya}}, \ and\ \bibinfo {author} {\bibfnamefont {J.~K.}\
		\bibnamefont {Kalita}},\ }\bibfield  {title} {\enquote {\bibinfo {title}
		{{Network Anomaly Detection: Methods, Systems and Tools}},}\ }\href {\doibase
	10.1109/SURV.2013.052213.00046} {\bibfield  {journal} {\bibinfo  {journal}
		{IEEE Communications Surveys {\&} Tutorials}\ }\textbf {\bibinfo {volume}
		{16}},\ \bibinfo {pages} {303--336} (\bibinfo {year} {2014})}\BibitemShut
{NoStop}%
\bibitem [{\citenamefont {Pimentel}\ \emph {et~al.}(2014)\citenamefont
	{Pimentel}, \citenamefont {Clifton}, \citenamefont {Clifton},\ and\
	\citenamefont {Tarassenko}}]{Pimentel2014}%
\BibitemOpen
\bibfield  {author} {\bibinfo {author} {\bibfnamefont {M.~A.}\ \bibnamefont
		{Pimentel}}, \bibinfo {author} {\bibfnamefont {D.~A.}\ \bibnamefont
		{Clifton}}, \bibinfo {author} {\bibfnamefont {L.}~\bibnamefont {Clifton}}, \
	and\ \bibinfo {author} {\bibfnamefont {L.}~\bibnamefont {Tarassenko}},\
}\bibfield  {title} {\enquote {\bibinfo {title} {{A review of novelty
			detection}},}\ }\href {\doibase 10.1016/j.sigpro.2013.12.026} {\bibfield
{journal} {\bibinfo  {journal} {Signal Processing}\ }\textbf {\bibinfo
	{volume} {99}},\ \bibinfo {pages} {215--249} (\bibinfo {year}
{2014})}\BibitemShut {NoStop}%
\bibitem [{\citenamefont {Libbrecht}\ and\ \citenamefont
	{Noble}(2015)}]{Libbrecht2015}%
\BibitemOpen
\bibfield  {author} {\bibinfo {author} {\bibfnamefont {M.~W.}\ \bibnamefont
		{Libbrecht}}\ and\ \bibinfo {author} {\bibfnamefont {W.~S.}\ \bibnamefont
		{Noble}},\ }\bibfield  {title} {\enquote {\bibinfo {title} {{Machine learning
				applications in genetics and genomics}},}\ }\href {\doibase 10.1038/nrg3920}
{\bibfield  {journal} {\bibinfo  {journal} {Nature Reviews Genetics}\
	}\textbf {\bibinfo {volume} {16}},\ \bibinfo {pages} {321--332} (\bibinfo
	{year} {2015})},\ \Eprint {http://arxiv.org/abs/15334406} {arXiv:15334406}
\BibitemShut {NoStop}%
\bibitem [{\citenamefont {Hammerschmidt}\ \emph {et~al.}(2016)\citenamefont
	{Hammerschmidt}, \citenamefont {Barth}, \citenamefont {Pomplun},
	\citenamefont {Burger}, \citenamefont {Becker},\ and\ \citenamefont
	{Schmidt}}]{Hammerschmidt2016}%
\BibitemOpen
\bibfield  {author} {\bibinfo {author} {\bibfnamefont {M.}~\bibnamefont
		{Hammerschmidt}}, \bibinfo {author} {\bibfnamefont {C.}~\bibnamefont
		{Barth}}, \bibinfo {author} {\bibfnamefont {J.}~\bibnamefont {Pomplun}},
	\bibinfo {author} {\bibfnamefont {S.}~\bibnamefont {Burger}}, \bibinfo
	{author} {\bibfnamefont {C.}~\bibnamefont {Becker}}, \ and\ \bibinfo {author}
	{\bibfnamefont {F.}~\bibnamefont {Schmidt}},\ }\bibfield  {title} {\enquote
	{\bibinfo {title} {{Reconstruction of photonic crystal geometries using a
				reduced basis method for nonlinear outputs}},}\ }\href {\doibase
	http://dx.doi.org/10.1117/12.2212482} {\bibfield  {journal} {\bibinfo
		{journal} {Proceedings of SPIE}\ }\textbf {\bibinfo {volume} {9756}},\
	\bibinfo {pages} {97561R} (\bibinfo {year} {2016})}\BibitemShut {NoStop}%
\bibitem [{\citenamefont {Smajic}, \citenamefont {Hafner},\ and\ \citenamefont
	{Erni}(2004)}]{Smajic2004}%
\BibitemOpen
\bibfield  {author} {\bibinfo {author} {\bibfnamefont {J.}~\bibnamefont
		{Smajic}}, \bibinfo {author} {\bibfnamefont {C.}~\bibnamefont {Hafner}}, \
	and\ \bibinfo {author} {\bibfnamefont {D.}~\bibnamefont {Erni}},\ }\bibfield
{title} {\enquote {\bibinfo {title} {{Optimization of photonic crystal
				structures}},}\ }\href {\doibase 10.1364/JOSAA.21.002223} {\bibfield
	{journal} {\bibinfo  {journal} {Journal of the Optical Society of America A}\
	}\textbf {\bibinfo {volume} {21}},\ \bibinfo {pages} {2223} (\bibinfo {year}
	{2004})}\BibitemShut {NoStop}%
\bibitem [{\citenamefont {Hakansson}, \citenamefont {Sanchez-Deh},\ and\
	\citenamefont {Sanchis}(2005)}]{Hakansson2005}%
\BibitemOpen
\bibfield  {author} {\bibinfo {author} {\bibfnamefont {A.}~\bibnamefont
		{Hakansson}}, \bibinfo {author} {\bibfnamefont {J.}~\bibnamefont
		{Sanchez-Deh}}, \ and\ \bibinfo {author} {\bibfnamefont {L.}~\bibnamefont
		{Sanchis}},\ }\bibfield  {title} {\enquote {\bibinfo {title} {{Inverse design
				of photonic crystal devices}},}\ }\href {\doibase 10.1109/JSAC.2005.851190}
{\bibfield  {journal} {\bibinfo  {journal} {IEEE Journal on Selected Areas in
			Communications}\ }\textbf {\bibinfo {volume} {23}},\ \bibinfo {pages}
	{1365--1371} (\bibinfo {year} {2005})},\ \Eprint
{http://arxiv.org/abs/0503529} {arXiv:0503529 [cond-mat]} \BibitemShut
{NoStop}%
\bibitem [{\citenamefont {Lu}(2013)}]{Lu2013}%
\BibitemOpen
\bibfield  {author} {\bibinfo {author} {\bibfnamefont {J.}~\bibnamefont
		{Lu}},\ }\emph {\bibinfo {title} {{Nanophotonic Computational Design}}},\
\href@noop {} {\bibinfo {type} {Dissertation}},\ \bibinfo  {school} {Stanford
	University} (\bibinfo {year} {2013})\BibitemShut {NoStop}%
\bibitem [{\citenamefont {Lu}\ and\ \citenamefont
	{Vu{\v{c}}kovi{\'{c}}}(2013)}]{Lu2013a}%
\BibitemOpen
\bibfield  {author} {\bibinfo {author} {\bibfnamefont {J.}~\bibnamefont
		{Lu}}\ and\ \bibinfo {author} {\bibfnamefont {J.}~\bibnamefont
		{Vu{\v{c}}kovi{\'{c}}}},\ }\bibfield  {title} {\enquote {\bibinfo {title}
		{{Nanophotonic computational design}},}\ }\href {\doibase
	10.1364/OE.21.013351} {\bibfield  {journal} {\bibinfo  {journal} {Optics
			Express}\ }\textbf {\bibinfo {volume} {21}},\ \bibinfo {pages} {13351}
	(\bibinfo {year} {2013})},\ \Eprint {http://arxiv.org/abs/1303.5823}
{arXiv:1303.5823} \BibitemShut {NoStop}%
\bibitem [{\citenamefont {Piggott}\ \emph {et~al.}(2015)\citenamefont
	{Piggott}, \citenamefont {Lu}, \citenamefont {Lagoudakis}, \citenamefont
	{Petykiewicz}, \citenamefont {Babinec},\ and\ \citenamefont
	{Vu{\v{c}}kovi{\'{c}}}}]{Piggott2015}%
\BibitemOpen
\bibfield  {author} {\bibinfo {author} {\bibfnamefont {A.~Y.}\ \bibnamefont
		{Piggott}}, \bibinfo {author} {\bibfnamefont {J.}~\bibnamefont {Lu}},
	\bibinfo {author} {\bibfnamefont {K.~G.}\ \bibnamefont {Lagoudakis}},
	\bibinfo {author} {\bibfnamefont {J.}~\bibnamefont {Petykiewicz}}, \bibinfo
	{author} {\bibfnamefont {T.~M.}\ \bibnamefont {Babinec}}, \ and\ \bibinfo
	{author} {\bibfnamefont {J.}~\bibnamefont {Vu{\v{c}}kovi{\'{c}}}},\
}\bibfield  {title} {\enquote {\bibinfo {title} {{Inverse design and
			demonstration of a compact and broadband on-chip wavelength
			demultiplexer}},}\ }\href {\doibase 10.1038/nphoton.2015.69} {\bibfield
{journal} {\bibinfo  {journal} {Nature Photonics}\ }\textbf {\bibinfo
	{volume} {9}},\ \bibinfo {pages} {374--377} (\bibinfo {year} {2015})},\
\Eprint {http://arxiv.org/abs/1504.00095} {arXiv:1504.00095} \BibitemShut
{NoStop}%
\bibitem [{\citenamefont {Rosenblatt}, \citenamefont {Sharon},\ and\
	\citenamefont {Friesem}(1997)}]{Rosenblatt1997}%
\BibitemOpen
\bibfield  {author} {\bibinfo {author} {\bibfnamefont {D.}~\bibnamefont
		{Rosenblatt}}, \bibinfo {author} {\bibfnamefont {A.}~\bibnamefont {Sharon}},
	\ and\ \bibinfo {author} {\bibfnamefont {A.}~\bibnamefont {Friesem}},\
}\bibfield  {title} {\enquote {\bibinfo {title} {{Resonant grating waveguide
			structures}},}\ }\href {\doibase 10.1109/3.641320} {\bibfield  {journal}
{\bibinfo  {journal} {IEEE Journal of Quantum Electronics}\ }\textbf
{\bibinfo {volume} {33}},\ \bibinfo {pages} {2038--2059} (\bibinfo {year}
{1997})}\BibitemShut {NoStop}%
\bibitem [{\citenamefont {Astratov}\ \emph
	{et~al.}(1999{\natexlab{a}})\citenamefont {Astratov}, \citenamefont
	{Whittaker}, \citenamefont {Culshaw}, \citenamefont {Stevenson},
	\citenamefont {Skolnick}, \citenamefont {Krauss},\ and\ \citenamefont {{De La
			Rue}}}]{Astratov1999}%
\BibitemOpen
\bibfield  {author} {\bibinfo {author} {\bibfnamefont {V.~N.}\ \bibnamefont
		{Astratov}}, \bibinfo {author} {\bibfnamefont {D.~M.}\ \bibnamefont
		{Whittaker}}, \bibinfo {author} {\bibfnamefont {I.~S.}\ \bibnamefont
		{Culshaw}}, \bibinfo {author} {\bibfnamefont {R.~M.}\ \bibnamefont
		{Stevenson}}, \bibinfo {author} {\bibfnamefont {M.~S.}\ \bibnamefont
		{Skolnick}}, \bibinfo {author} {\bibfnamefont {T.~F.}\ \bibnamefont
		{Krauss}}, \ and\ \bibinfo {author} {\bibfnamefont {R.~M.}\ \bibnamefont {{De
				La Rue}}},\ }\bibfield  {title} {\enquote {\bibinfo {title} {{Photonic
				band-structure effects in the reflectivity of periodically patterned
				waveguides}},}\ }\href {\doibase 10.1103/PhysRevB.60.R16255} {\bibfield
	{journal} {\bibinfo  {journal} {Physical Review B}\ }\textbf {\bibinfo
		{volume} {60}},\ \bibinfo {pages} {R16255--R16258} (\bibinfo {year}
	{1999}{\natexlab{a}})}\BibitemShut {NoStop}%
\bibitem [{\citenamefont {Astratov}\ \emph
	{et~al.}(1999{\natexlab{b}})\citenamefont {Astratov}, \citenamefont
	{Culshaw}, \citenamefont {Stevenson}, \citenamefont {Whittaker},
	\citenamefont {Skolnick}, \citenamefont {Krauss},\ and\ \citenamefont {{De La
			Rue}}}]{Astratov1999a}%
\BibitemOpen
\bibfield  {author} {\bibinfo {author} {\bibfnamefont {V.~N.}\ \bibnamefont
		{Astratov}}, \bibinfo {author} {\bibfnamefont {I.~S.}\ \bibnamefont
		{Culshaw}}, \bibinfo {author} {\bibfnamefont {R.~M.}\ \bibnamefont
		{Stevenson}}, \bibinfo {author} {\bibfnamefont {D.~M.}\ \bibnamefont
		{Whittaker}}, \bibinfo {author} {\bibfnamefont {M.~S.}\ \bibnamefont
		{Skolnick}}, \bibinfo {author} {\bibfnamefont {T.~F.}\ \bibnamefont
		{Krauss}}, \ and\ \bibinfo {author} {\bibfnamefont {R.~M.}\ \bibnamefont {{De
				La Rue}}},\ }\bibfield  {title} {\enquote {\bibinfo {title} {{Resonant
				coupling of near-infrared radiation to photonic band structure
				waveguides}},}\ }\href {\doibase 10.1109/50.802994} {\bibfield  {journal}
	{\bibinfo  {journal} {Journal of Lightwave Technology}\ }\textbf {\bibinfo
		{volume} {17}},\ \bibinfo {pages} {2050--2057} (\bibinfo {year}
	{1999}{\natexlab{b}})}\BibitemShut {NoStop}%
\bibitem [{\citenamefont {Erchak}\ \emph {et~al.}(2001)\citenamefont {Erchak},
	\citenamefont {Ripin}, \citenamefont {Fan}, \citenamefont {Rakich},
	\citenamefont {Joannopoulos}, \citenamefont {Ippen}, \citenamefont
	{Petrich},\ and\ \citenamefont {Kolodziejski}}]{Erchak2001}%
\BibitemOpen
\bibfield  {author} {\bibinfo {author} {\bibfnamefont {A.~A.}\ \bibnamefont
		{Erchak}}, \bibinfo {author} {\bibfnamefont {D.~J.}\ \bibnamefont {Ripin}},
	\bibinfo {author} {\bibfnamefont {S.}~\bibnamefont {Fan}}, \bibinfo {author}
	{\bibfnamefont {P.}~\bibnamefont {Rakich}}, \bibinfo {author} {\bibfnamefont
		{J.~D.}\ \bibnamefont {Joannopoulos}}, \bibinfo {author} {\bibfnamefont
		{E.~P.}\ \bibnamefont {Ippen}}, \bibinfo {author} {\bibfnamefont {G.~S.}\
		\bibnamefont {Petrich}}, \ and\ \bibinfo {author} {\bibfnamefont {L.~A.}\
		\bibnamefont {Kolodziejski}},\ }\bibfield  {title} {\enquote {\bibinfo
		{title} {{Enhanced coupling to vertical radiation using a two-dimensional
				photonic crystal in a semiconductor light-emitting diode}},}\ }\href
{\doibase 10.1063/1.1342048} {\bibfield  {journal} {\bibinfo  {journal}
		{Applied Physics Letters}\ }\textbf {\bibinfo {volume} {78}},\ \bibinfo
	{pages} {563--565} (\bibinfo {year} {2001})}\BibitemShut {NoStop}%
\bibitem [{\citenamefont {Ochiai}\ and\ \citenamefont
	{Sakoda}(2001)}]{Ochiai2001}%
\BibitemOpen
\bibfield  {author} {\bibinfo {author} {\bibfnamefont {T.}~\bibnamefont
		{Ochiai}}\ and\ \bibinfo {author} {\bibfnamefont {K.}~\bibnamefont
		{Sakoda}},\ }\bibfield  {title} {\enquote {\bibinfo {title} {{Dispersion
				relation and optical transmittance of a hexagonal photonic crystal slab}},}\
}\href {\doibase 10.1103/PhysRevB.63.125107} {\bibfield  {journal} {\bibinfo
	{journal} {Physical Review B}\ }\textbf {\bibinfo {volume} {63}},\ \bibinfo
{pages} {125107} (\bibinfo {year} {2001})}\BibitemShut {NoStop}%
\bibitem [{\citenamefont {Chutinan}\ and\ \citenamefont
	{John}(2008)}]{Chutinan2008}%
\BibitemOpen
\bibfield  {author} {\bibinfo {author} {\bibfnamefont {A.}~\bibnamefont
		{Chutinan}}\ and\ \bibinfo {author} {\bibfnamefont {S.}~\bibnamefont
		{John}},\ }\bibfield  {title} {\enquote {\bibinfo {title} {{Light trapping
				and absorption optimization in certain thin-film photonic crystal
				architectures}},}\ }\href {\doibase 10.1103/PhysRevA.78.023825} {\bibfield
	{journal} {\bibinfo  {journal} {Physical Review A}\ }\textbf {\bibinfo
		{volume} {78}},\ \bibinfo {pages} {023825} (\bibinfo {year}
	{2008})}\BibitemShut {NoStop}%
\bibitem [{\citenamefont {Han}\ and\ \citenamefont {Chen}(2010)}]{Han2010a}%
\BibitemOpen
\bibfield  {author} {\bibinfo {author} {\bibfnamefont {S.~E.}\ \bibnamefont
		{Han}}\ and\ \bibinfo {author} {\bibfnamefont {G.}~\bibnamefont {Chen}},\
}\bibfield  {title} {\enquote {\bibinfo {title} {{Toward the Lambertian limit
			of light trapping in thin nanostructured silicon solar cells.}}}\ }\href
{\doibase 10.1021/nl1029804} {\bibfield  {journal} {\bibinfo  {journal} {Nano
			letters}\ }\textbf {\bibinfo {volume} {10}},\ \bibinfo {pages} {4692--6}
	(\bibinfo {year} {2010})}\BibitemShut {NoStop}%
\bibitem [{\citenamefont {John}(2012)}]{John2012}%
\BibitemOpen
\bibfield  {author} {\bibinfo {author} {\bibfnamefont {S.}~\bibnamefont
		{John}},\ }\bibfield  {title} {\enquote {\bibinfo {title} {{Why trap
				light?}}}\ }\href {\doibase 10.1038/nmat3503} {\bibfield  {journal} {\bibinfo
		{journal} {Nature materials}\ }\textbf {\bibinfo {volume} {11}},\ \bibinfo
	{pages} {997--9} (\bibinfo {year} {2012})}\BibitemShut {NoStop}%
\bibitem [{\citenamefont {Mellor}\ \emph {et~al.}(2013)\citenamefont {Mellor},
	\citenamefont {Hauser}, \citenamefont {Wellens}, \citenamefont {Benick},
	\citenamefont {Eisenlohr}, \citenamefont {Peters}, \citenamefont {Guttowski},
	\citenamefont {Tob{\'{i}}as}, \citenamefont {Mart{\'{i}}}, \citenamefont
	{Luque},\ and\ \citenamefont {Bl{\"{a}}si}}]{Mellor2013}%
\BibitemOpen
\bibfield  {author} {\bibinfo {author} {\bibfnamefont {A.}~\bibnamefont
		{Mellor}}, \bibinfo {author} {\bibfnamefont {H.}~\bibnamefont {Hauser}},
	\bibinfo {author} {\bibfnamefont {C.}~\bibnamefont {Wellens}}, \bibinfo
	{author} {\bibfnamefont {J.}~\bibnamefont {Benick}}, \bibinfo {author}
	{\bibfnamefont {J.}~\bibnamefont {Eisenlohr}}, \bibinfo {author}
	{\bibfnamefont {M.}~\bibnamefont {Peters}}, \bibinfo {author} {\bibfnamefont
		{A.}~\bibnamefont {Guttowski}}, \bibinfo {author} {\bibfnamefont
		{I.}~\bibnamefont {Tob{\'{i}}as}}, \bibinfo {author} {\bibfnamefont
		{A.}~\bibnamefont {Mart{\'{i}}}}, \bibinfo {author} {\bibfnamefont
		{A.}~\bibnamefont {Luque}}, \ and\ \bibinfo {author} {\bibfnamefont
		{B.}~\bibnamefont {Bl{\"{a}}si}},\ }\bibfield  {title} {\enquote {\bibinfo
		{title} {{Nanoimprinted diffraction gratings for crystalline silicon solar
				cells: implementation, characterization and simulation.}}}\ }\href
{http://www.ncbi.nlm.nih.gov/pubmed/23482292} {\bibfield  {journal} {\bibinfo
		{journal} {Optics express}\ }\textbf {\bibinfo {volume} {21 Suppl 2}},\
	\bibinfo {pages} {A295--304} (\bibinfo {year} {2013})}\BibitemShut {NoStop}%
\bibitem [{\citenamefont {Branham}\ \emph {et~al.}(2015)\citenamefont
	{Branham}, \citenamefont {Hsu}, \citenamefont {Yerci}, \citenamefont
	{Loomis}, \citenamefont {Boriskina}, \citenamefont {Hoard}, \citenamefont
	{Han},\ and\ \citenamefont {Chen}}]{Branham2015}%
\BibitemOpen
\bibfield  {author} {\bibinfo {author} {\bibfnamefont {M.~S.}\ \bibnamefont
		{Branham}}, \bibinfo {author} {\bibfnamefont {W.-C.}\ \bibnamefont {Hsu}},
	\bibinfo {author} {\bibfnamefont {S.}~\bibnamefont {Yerci}}, \bibinfo
	{author} {\bibfnamefont {J.}~\bibnamefont {Loomis}}, \bibinfo {author}
	{\bibfnamefont {S.~V.}\ \bibnamefont {Boriskina}}, \bibinfo {author}
	{\bibfnamefont {B.~R.}\ \bibnamefont {Hoard}}, \bibinfo {author}
	{\bibfnamefont {S.~E.}\ \bibnamefont {Han}}, \ and\ \bibinfo {author}
	{\bibfnamefont {G.}~\bibnamefont {Chen}},\ }\bibfield  {title} {\enquote
	{\bibinfo {title} {{15.7{\%} Efficient 10-$\mu$m-Thick Crystalline Silicon
				Solar Cells Using Periodic Nanostructures}},}\ }\href {\doibase
	10.1002/adma.201405511} {\bibfield  {journal} {\bibinfo  {journal} {Advanced
			Materials}\ }\textbf {\bibinfo {volume} {27}},\ \bibinfo {pages} {2182--2188}
	(\bibinfo {year} {2015})}\BibitemShut {NoStop}%
\bibitem [{\citenamefont {Fan}\ \emph {et~al.}(1997)\citenamefont {Fan},
	\citenamefont {Villeneuve}, \citenamefont {Joannopoulos},\ and\ \citenamefont
	{Schubert}}]{Fan1997}%
\BibitemOpen
\bibfield  {author} {\bibinfo {author} {\bibfnamefont {S.}~\bibnamefont
		{Fan}}, \bibinfo {author} {\bibfnamefont {P.~R.}\ \bibnamefont {Villeneuve}},
	\bibinfo {author} {\bibfnamefont {J.~D.}\ \bibnamefont {Joannopoulos}}, \
	and\ \bibinfo {author} {\bibfnamefont {E.~F.}\ \bibnamefont {Schubert}},\
}\bibfield  {title} {\enquote {\bibinfo {title} {{High Extraction Efficiency
			of Spontaneous Emission from Slabs of Photonic Crystals}},}\ }\href {\doibase
10.1103/PhysRevLett.78.3294} {\bibfield  {journal} {\bibinfo  {journal}
	{Physical Review Letters}\ }\textbf {\bibinfo {volume} {78}},\ \bibinfo
{pages} {3294--3297} (\bibinfo {year} {1997})}\BibitemShut {NoStop}%
\bibitem [{\citenamefont {Wiesmann}\ \emph {et~al.}(2009)\citenamefont
	{Wiesmann}, \citenamefont {Bergenek}, \citenamefont {Linder},\ and\
	\citenamefont {Schwarz}}]{Wiesmann2009}%
\BibitemOpen
\bibfield  {author} {\bibinfo {author} {\bibfnamefont {C.}~\bibnamefont
		{Wiesmann}}, \bibinfo {author} {\bibfnamefont {K.}~\bibnamefont {Bergenek}},
	\bibinfo {author} {\bibfnamefont {N.}~\bibnamefont {Linder}}, \ and\ \bibinfo
	{author} {\bibfnamefont {U.}~\bibnamefont {Schwarz}},\ }\bibfield  {title}
{\enquote {\bibinfo {title} {{Photonic crystal LEDs - designing light
				extraction}},}\ }\href {\doibase 10.1002/lpor.200810053} {\bibfield
	{journal} {\bibinfo  {journal} {Laser {\&} Photonics Review}\ }\textbf
	{\bibinfo {volume} {3}},\ \bibinfo {pages} {262--286} (\bibinfo {year}
	{2009})}\BibitemShut {NoStop}%
\bibitem [{\citenamefont {Cunningham}\ \emph {et~al.}(2016)\citenamefont
	{Cunningham}, \citenamefont {Zhang}, \citenamefont {Zhuo}, \citenamefont
	{Kwon},\ and\ \citenamefont {Race}}]{Cunningham2016}%
\BibitemOpen
\bibfield  {author} {\bibinfo {author} {\bibfnamefont {B.~T.}\ \bibnamefont
		{Cunningham}}, \bibinfo {author} {\bibfnamefont {M.}~\bibnamefont {Zhang}},
	\bibinfo {author} {\bibfnamefont {Y.}~\bibnamefont {Zhuo}}, \bibinfo {author}
	{\bibfnamefont {L.}~\bibnamefont {Kwon}}, \ and\ \bibinfo {author}
	{\bibfnamefont {C.}~\bibnamefont {Race}},\ }\bibfield  {title} {\enquote
	{\bibinfo {title} {{Recent Advances in Biosensing With Photonic Crystal
				Surfaces: A Review}},}\ }\href {\doibase 10.1109/JSEN.2015.2429738}
{\bibfield  {journal} {\bibinfo  {journal} {IEEE Sensors Journal}\ }\textbf
	{\bibinfo {volume} {16}},\ \bibinfo {pages} {3349--3366} (\bibinfo {year}
	{2016})}\BibitemShut {NoStop}%
\bibitem [{\citenamefont {Block}\ \emph {et~al.}(2009)\citenamefont {Block},
	\citenamefont {Mathias}, \citenamefont {Ganesh}, \citenamefont {Jones},
	\citenamefont {Dorvel}, \citenamefont {Chaudhery}, \citenamefont {Vodkin},
	\citenamefont {Bashir},\ and\ \citenamefont {Cunningham}}]{Block2009}%
\BibitemOpen
\bibfield  {author} {\bibinfo {author} {\bibfnamefont {I.~D.}\ \bibnamefont
		{Block}}, \bibinfo {author} {\bibfnamefont {P.~C.}\ \bibnamefont {Mathias}},
	\bibinfo {author} {\bibfnamefont {N.}~\bibnamefont {Ganesh}}, \bibinfo
	{author} {\bibfnamefont {S.~I.}\ \bibnamefont {Jones}}, \bibinfo {author}
	{\bibfnamefont {B.~R.}\ \bibnamefont {Dorvel}}, \bibinfo {author}
	{\bibfnamefont {V.}~\bibnamefont {Chaudhery}}, \bibinfo {author}
	{\bibfnamefont {L.~O.}\ \bibnamefont {Vodkin}}, \bibinfo {author}
	{\bibfnamefont {R.}~\bibnamefont {Bashir}}, \ and\ \bibinfo {author}
	{\bibfnamefont {B.~T.}\ \bibnamefont {Cunningham}},\ }\bibfield  {title}
{\enquote {\bibinfo {title} {{A detection instrument for
				enhanced-fluorescence and label-free imaging on photonic crystal
				surfaces}},}\ }\href {\doibase 10.1364/OE.17.013222} {\bibfield  {journal}
	{\bibinfo  {journal} {Optics Express}\ }\textbf {\bibinfo {volume} {17}},\
	\bibinfo {pages} {13222} (\bibinfo {year} {2009})}\BibitemShut {NoStop}%
\bibitem [{\citenamefont {Ganesh}\ \emph
	{et~al.}(2008{\natexlab{a}})\citenamefont {Ganesh}, \citenamefont {Mathias},
	\citenamefont {Zhang},\ and\ \citenamefont {Cunningham}}]{Ganesh2008a}%
\BibitemOpen
\bibfield  {author} {\bibinfo {author} {\bibfnamefont {N.}~\bibnamefont
		{Ganesh}}, \bibinfo {author} {\bibfnamefont {P.~C.}\ \bibnamefont {Mathias}},
	\bibinfo {author} {\bibfnamefont {W.}~\bibnamefont {Zhang}}, \ and\ \bibinfo
	{author} {\bibfnamefont {B.~T.}\ \bibnamefont {Cunningham}},\ }\bibfield
{title} {\enquote {\bibinfo {title} {{Distance dependence of fluorescence
				enhancement from photonic crystal surfaces}},}\ }\href {\doibase
	10.1063/1.2906175} {\bibfield  {journal} {\bibinfo  {journal} {Journal of
			Applied Physics}\ }\textbf {\bibinfo {volume} {103}},\ \bibinfo {pages}
	{083104} (\bibinfo {year} {2008}{\natexlab{a}})}\BibitemShut {NoStop}%
\bibitem [{\citenamefont {Threm}, \citenamefont {Nazirizadeh},\ and\
	\citenamefont {Gerken}(2012)}]{Threm2012}%
\BibitemOpen
\bibfield  {author} {\bibinfo {author} {\bibfnamefont {D.}~\bibnamefont
		{Threm}}, \bibinfo {author} {\bibfnamefont {Y.}~\bibnamefont {Nazirizadeh}},
	\ and\ \bibinfo {author} {\bibfnamefont {M.}~\bibnamefont {Gerken}},\
}\bibfield  {title} {\enquote {\bibinfo {title} {{Photonic crystal biosensors
			towards on-chip integration}},}\ }\href {\doibase 10.1002/jbio.201200039}
{\bibfield  {journal} {\bibinfo  {journal} {Journal of Biophotonics}\
	}\textbf {\bibinfo {volume} {5}},\ \bibinfo {pages} {601--616} (\bibinfo
	{year} {2012})}\BibitemShut {NoStop}%
\bibitem [{\citenamefont {Boroditsky}\ \emph {et~al.}(1999)\citenamefont
	{Boroditsky}, \citenamefont {Vrijen}, \citenamefont {Krauss}, \citenamefont
	{Coccioli}, \citenamefont {Bhat},\ and\ \citenamefont
	{Yablonovitch}}]{Boroditsky1999}%
\BibitemOpen
\bibfield  {author} {\bibinfo {author} {\bibfnamefont {M.}~\bibnamefont
		{Boroditsky}}, \bibinfo {author} {\bibfnamefont {R.}~\bibnamefont {Vrijen}},
	\bibinfo {author} {\bibfnamefont {T.}~\bibnamefont {Krauss}}, \bibinfo
	{author} {\bibfnamefont {R.}~\bibnamefont {Coccioli}}, \bibinfo {author}
	{\bibfnamefont {R.}~\bibnamefont {Bhat}}, \ and\ \bibinfo {author}
	{\bibfnamefont {E.}~\bibnamefont {Yablonovitch}},\ }\bibfield  {title}
{\enquote {\bibinfo {title} {{Spontaneous emission extraction and Purcell
				enhancement from thin-film 2-D photonic crystals}},}\ }\href {\doibase
	10.1109/50.803000} {\bibfield  {journal} {\bibinfo  {journal} {Journal of
			Lightwave Technology}\ }\textbf {\bibinfo {volume} {17}},\ \bibinfo {pages}
	{2096--2112} (\bibinfo {year} {1999})}\BibitemShut {NoStop}%
\bibitem [{\citenamefont {Ganesh}\ \emph
	{et~al.}(2008{\natexlab{b}})\citenamefont {Ganesh}, \citenamefont {Block},
	\citenamefont {Mathias}, \citenamefont {Zhang}, \citenamefont {Chow},
	\citenamefont {Malyarchuk},\ and\ \citenamefont {Cunningham}}]{Ganesh2008}%
\BibitemOpen
\bibfield  {author} {\bibinfo {author} {\bibfnamefont {N.}~\bibnamefont
		{Ganesh}}, \bibinfo {author} {\bibfnamefont {I.~D.}\ \bibnamefont {Block}},
	\bibinfo {author} {\bibfnamefont {P.~C.}\ \bibnamefont {Mathias}}, \bibinfo
	{author} {\bibfnamefont {W.}~\bibnamefont {Zhang}}, \bibinfo {author}
	{\bibfnamefont {E.}~\bibnamefont {Chow}}, \bibinfo {author} {\bibfnamefont
		{V.}~\bibnamefont {Malyarchuk}}, \ and\ \bibinfo {author} {\bibfnamefont
		{B.~T.}\ \bibnamefont {Cunningham}},\ }\bibfield  {title} {\enquote {\bibinfo
		{title} {{Leaky-mode assisted fluorescence extraction: application to
				fluorescence enhancement biosensors.}}}\ }\href {\doibase
	10.1364/OE.16.021626} {\bibfield  {journal} {\bibinfo  {journal} {Optics
			express}\ }\textbf {\bibinfo {volume} {16}},\ \bibinfo {pages} {21626--21640}
	(\bibinfo {year} {2008}{\natexlab{b}})}\BibitemShut {NoStop}%
\bibitem [{\citenamefont {Ondi{\v{c}}}\ \emph {et~al.}(2011)\citenamefont
	{Ondi{\v{c}}}, \citenamefont {Dohnalová}, \citenamefont {Ledinský},
	\citenamefont {Kromka}, \citenamefont {Babchenko},\ and\ \citenamefont
	{Rezek}}]{Ondic2011a}%
\BibitemOpen
\bibfield  {author} {\bibinfo {author} {\bibfnamefont {L.}~\bibnamefont
		{Ondi{\v{c}}}}, \bibinfo {author} {\bibfnamefont {K.}~\bibnamefont
		{Dohnalová}}, \bibinfo {author} {\bibfnamefont {M.}~\bibnamefont
		{Ledinský}}, \bibinfo {author} {\bibfnamefont {A.}~\bibnamefont {Kromka}},
	\bibinfo {author} {\bibfnamefont {O.}~\bibnamefont {Babchenko}}, \ and\
	\bibinfo {author} {\bibfnamefont {B.}~\bibnamefont {Rezek}},\ }\bibfield
{title} {\enquote {\bibinfo {title} {{Effective Extraction of
				Photoluminescence from a Diamond Layer with a Photonic Crystal}},}\ }\href
{\doibase 10.1021/nn1021555} {\bibfield  {journal} {\bibinfo  {journal} {ACS
			Nano}\ }\textbf {\bibinfo {volume} {5}},\ \bibinfo {pages} {346--350}
	(\bibinfo {year} {2011})}\BibitemShut {NoStop}%
\bibitem [{\citenamefont {Ondi{\v{c}}}\ \emph {et~al.}(2012)\citenamefont
	{Ondi{\v{c}}}, \citenamefont {Babchenko}, \citenamefont {Varga},
	\citenamefont {Kromka}, \citenamefont {{\v{C}}tyrok{\'{y}}},\ and\
	\citenamefont {Pelant}}]{Ondic2012}%
\BibitemOpen
\bibfield  {author} {\bibinfo {author} {\bibfnamefont {L.}~\bibnamefont
		{Ondi{\v{c}}}}, \bibinfo {author} {\bibfnamefont {O.}~\bibnamefont
		{Babchenko}}, \bibinfo {author} {\bibfnamefont {M.}~\bibnamefont {Varga}},
	\bibinfo {author} {\bibfnamefont {A.}~\bibnamefont {Kromka}}, \bibinfo
	{author} {\bibfnamefont {J.}~\bibnamefont {{\v{C}}tyrok{\'{y}}}}, \ and\
	\bibinfo {author} {\bibfnamefont {I.}~\bibnamefont {Pelant}},\ }\bibfield
{title} {\enquote {\bibinfo {title} {{Diamond photonic crystal slab: Leaky
				modes and modified photoluminescence emission of surface-deposited quantum
				dots}},}\ }\href {\doibase 10.1038/srep00914} {\bibfield  {journal} {\bibinfo
		{journal} {Scientific Reports}\ }\textbf {\bibinfo {volume} {2}},\ \bibinfo
	{pages} {914} (\bibinfo {year} {2012})}\BibitemShut {NoStop}%
\bibitem [{\citenamefont {Ondi{\v{c}}}\ \emph {et~al.}(2013)\citenamefont
	{Ondi{\v{c}}}, \citenamefont {Varga}, \citenamefont {Hru{\v{s}}ka},
	\citenamefont {Kromka}, \citenamefont {Herynkov{\'{a}}}, \citenamefont
	{H{\"{o}}nerlage},\ and\ \citenamefont {Pelant}}]{Ondic2013}%
\BibitemOpen
\bibfield  {author} {\bibinfo {author} {\bibfnamefont {L.}~\bibnamefont
		{Ondi{\v{c}}}}, \bibinfo {author} {\bibfnamefont {M.}~\bibnamefont {Varga}},
	\bibinfo {author} {\bibfnamefont {K.}~\bibnamefont {Hru{\v{s}}ka}}, \bibinfo
	{author} {\bibfnamefont {A.}~\bibnamefont {Kromka}}, \bibinfo {author}
	{\bibfnamefont {K.}~\bibnamefont {Herynkov{\'{a}}}}, \bibinfo {author}
	{\bibfnamefont {B.}~\bibnamefont {H{\"{o}}nerlage}}, \ and\ \bibinfo {author}
	{\bibfnamefont {I.}~\bibnamefont {Pelant}},\ }\bibfield  {title} {\enquote
	{\bibinfo {title} {{Two-dimensional photonic crystal slab with embedded
				silicon nanocrystals: Efficient photoluminescence extraction}},}\ }\href
{\doibase 10.1063/1.4812477} {\bibfield  {journal} {\bibinfo  {journal}
		{Applied Physics Letters}\ }\textbf {\bibinfo {volume} {102}},\ \bibinfo
	{pages} {251111} (\bibinfo {year} {2013})}\BibitemShut {NoStop}%
\bibitem [{\citenamefont {Adachi}\ \emph {et~al.}(2013)\citenamefont {Adachi},
	\citenamefont {Labelle}, \citenamefont {Thon}, \citenamefont {Lan},
	\citenamefont {Hoogland},\ and\ \citenamefont {Sargent}}]{Adachi2013}%
\BibitemOpen
\bibfield  {author} {\bibinfo {author} {\bibfnamefont {M.~M.}\ \bibnamefont
		{Adachi}}, \bibinfo {author} {\bibfnamefont {A.~J.}\ \bibnamefont {Labelle}},
	\bibinfo {author} {\bibfnamefont {S.~M.}\ \bibnamefont {Thon}}, \bibinfo
	{author} {\bibfnamefont {X.}~\bibnamefont {Lan}}, \bibinfo {author}
	{\bibfnamefont {S.}~\bibnamefont {Hoogland}}, \ and\ \bibinfo {author}
	{\bibfnamefont {E.~H.}\ \bibnamefont {Sargent}},\ }\bibfield  {title}
{\enquote {\bibinfo {title} {{Broadband solar absorption enhancement via
				periodic nanostructuring of electrodes}},}\ }\href {\doibase
	10.1038/srep02928} {\bibfield  {journal} {\bibinfo  {journal} {Scientific
			Reports}\ }\textbf {\bibinfo {volume} {3}},\ \bibinfo {pages} {2928}
	(\bibinfo {year} {2013})}\BibitemShut {NoStop}%
\bibitem [{\citenamefont {Kim}\ \emph {et~al.}(2013)\citenamefont {Kim},
	\citenamefont {Kim}, \citenamefont {Gao}, \citenamefont {Song}, \citenamefont
	{An}, \citenamefont {You}, \citenamefont {Lee}, \citenamefont {Jeong},
	\citenamefont {Lee}, \citenamefont {Jeong}, \citenamefont {Beard},\ and\
	\citenamefont {Jeong}}]{Kim2013}%
\BibitemOpen
\bibfield  {author} {\bibinfo {author} {\bibfnamefont {S.}~\bibnamefont
		{Kim}}, \bibinfo {author} {\bibfnamefont {J.~K.}\ \bibnamefont {Kim}},
	\bibinfo {author} {\bibfnamefont {J.}~\bibnamefont {Gao}}, \bibinfo {author}
	{\bibfnamefont {J.~H.}\ \bibnamefont {Song}}, \bibinfo {author}
	{\bibfnamefont {H.~J.}\ \bibnamefont {An}}, \bibinfo {author} {\bibfnamefont
		{T.-s.}\ \bibnamefont {You}}, \bibinfo {author} {\bibfnamefont {T.-S.}\
		\bibnamefont {Lee}}, \bibinfo {author} {\bibfnamefont {J.-R.}\ \bibnamefont
		{Jeong}}, \bibinfo {author} {\bibfnamefont {E.-S.}\ \bibnamefont {Lee}},
	\bibinfo {author} {\bibfnamefont {J.-H.}\ \bibnamefont {Jeong}}, \bibinfo
	{author} {\bibfnamefont {M.~C.}\ \bibnamefont {Beard}}, \ and\ \bibinfo
	{author} {\bibfnamefont {S.}~\bibnamefont {Jeong}},\ }\bibfield  {title}
{\enquote {\bibinfo {title} {{Lead Sulfide Nanocrystal Quantum Dot Solar
				Cells with Trenched ZnO Fabricated via Nanoimprinting}},}\ }\href {\doibase
	10.1021/am400443w} {\bibfield  {journal} {\bibinfo  {journal} {ACS Applied
			Materials {\&} Interfaces}\ }\textbf {\bibinfo {volume} {5}},\ \bibinfo
	{pages} {3803--3808} (\bibinfo {year} {2013})}\BibitemShut {NoStop}%
\bibitem [{\citenamefont {Su}\ \emph {et~al.}(2013)\citenamefont {Su},
	\citenamefont {Karuturi}, \citenamefont {Luo}, \citenamefont {Liu},
	\citenamefont {Liu}, \citenamefont {Guo}, \citenamefont {Sum}, \citenamefont
	{Deng}, \citenamefont {Fan}, \citenamefont {Liu},\ and\ \citenamefont
	{Tok}}]{Liu2013}%
\BibitemOpen
\bibfield  {author} {\bibinfo {author} {\bibfnamefont {L.~T.}\ \bibnamefont
		{Su}}, \bibinfo {author} {\bibfnamefont {S.~K.}\ \bibnamefont {Karuturi}},
	\bibinfo {author} {\bibfnamefont {J.}~\bibnamefont {Luo}}, \bibinfo {author}
	{\bibfnamefont {L.}~\bibnamefont {Liu}}, \bibinfo {author} {\bibfnamefont
		{X.}~\bibnamefont {Liu}}, \bibinfo {author} {\bibfnamefont {J.}~\bibnamefont
		{Guo}}, \bibinfo {author} {\bibfnamefont {T.~C.}\ \bibnamefont {Sum}},
	\bibinfo {author} {\bibfnamefont {R.}~\bibnamefont {Deng}}, \bibinfo {author}
	{\bibfnamefont {H.~J.}\ \bibnamefont {Fan}}, \bibinfo {author} {\bibfnamefont
		{X.}~\bibnamefont {Liu}}, \ and\ \bibinfo {author} {\bibfnamefont {A.~I.~Y.}\
		\bibnamefont {Tok}},\ }\bibfield  {title} {\enquote {\bibinfo {title}
		{{Photon Upconversion in Hetero-nanostructured Photoanodes for Enhanced
				Near-Infrared Light Harvesting}},}\ }\href {\doibase 10.1002/adma.201204353}
{\bibfield  {journal} {\bibinfo  {journal} {Advanced Materials}\ }\textbf
	{\bibinfo {volume} {25}},\ \bibinfo {pages} {1603--1607} (\bibinfo {year}
	{2013})}\BibitemShut {NoStop}%
\bibitem [{\citenamefont {Zhang}\ \emph {et~al.}(2010)\citenamefont {Zhang},
	\citenamefont {Deng}, \citenamefont {Shi}, \citenamefont {Zhang},\ and\
	\citenamefont {Zhao}}]{Zhang2010a}%
\BibitemOpen
\bibfield  {author} {\bibinfo {author} {\bibfnamefont {F.}~\bibnamefont
		{Zhang}}, \bibinfo {author} {\bibfnamefont {Y.}~\bibnamefont {Deng}},
	\bibinfo {author} {\bibfnamefont {Y.}~\bibnamefont {Shi}}, \bibinfo {author}
	{\bibfnamefont {R.}~\bibnamefont {Zhang}}, \ and\ \bibinfo {author}
	{\bibfnamefont {D.}~\bibnamefont {Zhao}},\ }\bibfield  {title} {\enquote
	{\bibinfo {title} {{Photoluminescence modification in upconversion rare-earth
				fluoride nanocrystal array constructed photonic crystals}},}\ }\href
{\doibase 10.1039/c000379d} {\bibfield  {journal} {\bibinfo  {journal}
		{Journal of Materials Chemistry}\ }\textbf {\bibinfo {volume} {20}},\
	\bibinfo {pages} {3895} (\bibinfo {year} {2010})}\BibitemShut {NoStop}%
\bibitem [{\citenamefont {Hofmann}\ \emph {et~al.}(2016)\citenamefont
	{Hofmann}, \citenamefont {Herter}, \citenamefont {Fischer}, \citenamefont
	{Gutmann},\ and\ \citenamefont {Goldschmidt}}]{Hofmann2016}%
\BibitemOpen
\bibfield  {author} {\bibinfo {author} {\bibfnamefont {C.~L.~M.}\
		\bibnamefont {Hofmann}}, \bibinfo {author} {\bibfnamefont {B.}~\bibnamefont
		{Herter}}, \bibinfo {author} {\bibfnamefont {S.}~\bibnamefont {Fischer}},
	\bibinfo {author} {\bibfnamefont {J.}~\bibnamefont {Gutmann}}, \ and\
	\bibinfo {author} {\bibfnamefont {J.~C.}\ \bibnamefont {Goldschmidt}},\
}\bibfield  {title} {\enquote {\bibinfo {title} {{Upconversion in a Bragg
			structure: photonic effects of a modified local density of states and
			irradiance on luminescence and upconversion quantum yield}},}\ }\href
{\doibase 10.1364/OE.24.014895} {\bibfield  {journal} {\bibinfo  {journal}
		{Optics Express}\ }\textbf {\bibinfo {volume} {24}},\ \bibinfo {pages}
	{14895} (\bibinfo {year} {2016})}\BibitemShut {NoStop}%
\bibitem [{\citenamefont {Ganesh}\ \emph {et~al.}(2007)\citenamefont {Ganesh},
	\citenamefont {Zhang}, \citenamefont {Mathias}, \citenamefont {Chow},
	\citenamefont {Soares}, \citenamefont {Malyarchuk}, \citenamefont {Smith},\
	and\ \citenamefont {Cunningham}}]{Ganesh2007}%
\BibitemOpen
\bibfield  {author} {\bibinfo {author} {\bibfnamefont {N.}~\bibnamefont
		{Ganesh}}, \bibinfo {author} {\bibfnamefont {W.}~\bibnamefont {Zhang}},
	\bibinfo {author} {\bibfnamefont {P.~C.}\ \bibnamefont {Mathias}}, \bibinfo
	{author} {\bibfnamefont {E.}~\bibnamefont {Chow}}, \bibinfo {author}
	{\bibfnamefont {J.~a. N.~T.}\ \bibnamefont {Soares}}, \bibinfo {author}
	{\bibfnamefont {V.}~\bibnamefont {Malyarchuk}}, \bibinfo {author}
	{\bibfnamefont {A.~D.}\ \bibnamefont {Smith}}, \ and\ \bibinfo {author}
	{\bibfnamefont {B.~T.}\ \bibnamefont {Cunningham}},\ }\bibfield  {title}
{\enquote {\bibinfo {title} {{Enhanced fluorescence emission from quantum
				dots on a photonic crystal surface}},}\ }\href {\doibase
	10.1038/nnano.2007.216} {\bibfield  {journal} {\bibinfo  {journal} {Nature
			Nanotechnology}\ }\textbf {\bibinfo {volume} {2}},\ \bibinfo {pages}
	{515--520} (\bibinfo {year} {2007})}\BibitemShut {NoStop}%
\bibitem [{\citenamefont {Barth}\ \emph {et~al.}(2017)\citenamefont {Barth},
	\citenamefont {Roder}, \citenamefont {Brodoceanu}, \citenamefont {Kraus},
	\citenamefont {Hammerschmidt}, \citenamefont {Burger},\ and\ \citenamefont
	{Becker}}]{Barth2017}%
\BibitemOpen
\bibfield  {author} {\bibinfo {author} {\bibfnamefont {C.}~\bibnamefont
		{Barth}}, \bibinfo {author} {\bibfnamefont {S.}~\bibnamefont {Roder}},
	\bibinfo {author} {\bibfnamefont {D.}~\bibnamefont {Brodoceanu}}, \bibinfo
	{author} {\bibfnamefont {T.}~\bibnamefont {Kraus}}, \bibinfo {author}
	{\bibfnamefont {M.}~\bibnamefont {Hammerschmidt}}, \bibinfo {author}
	{\bibfnamefont {S.}~\bibnamefont {Burger}}, \ and\ \bibinfo {author}
	{\bibfnamefont {C.}~\bibnamefont {Becker}},\ }\bibfield  {title} {\enquote
	{\bibinfo {title} {{Increased fluorescence of PbS quantum dots in photonic
				crystals by excitation enhancement}},}\ }\href {\doibase 10.1063/1.4995229}
{\bibfield  {journal} {\bibinfo  {journal} {Applied Physics Letters}\
	}\textbf {\bibinfo {volume} {111}},\ \bibinfo {pages} {031111} (\bibinfo
	{year} {2017})}\BibitemShut {NoStop}%
\bibitem [{\citenamefont {Becker}\ \emph {et~al.}(2014)\citenamefont {Becker},
	\citenamefont {Wyss}, \citenamefont {Eisenhauer}, \citenamefont {Probst},
	\citenamefont {Preidel}, \citenamefont {Hammerschmidt},\ and\ \citenamefont
	{Burger}}]{Becker2014a}%
\BibitemOpen
\bibfield  {author} {\bibinfo {author} {\bibfnamefont {C.}~\bibnamefont
		{Becker}}, \bibinfo {author} {\bibfnamefont {P.}~\bibnamefont {Wyss}},
	\bibinfo {author} {\bibfnamefont {D.}~\bibnamefont {Eisenhauer}}, \bibinfo
	{author} {\bibfnamefont {J.}~\bibnamefont {Probst}}, \bibinfo {author}
	{\bibfnamefont {V.}~\bibnamefont {Preidel}}, \bibinfo {author} {\bibfnamefont
		{M.}~\bibnamefont {Hammerschmidt}}, \ and\ \bibinfo {author} {\bibfnamefont
		{S.}~\bibnamefont {Burger}},\ }\bibfield  {title} {\enquote {\bibinfo {title}
		{5 × 5 cm² silicon photonic crystal slabs on glass and plastic foil
			exhibiting broadband absorption and high-intensity near-fields},}\ }\href
{\doibase 10.1038/srep05886} {\bibfield  {journal} {\bibinfo  {journal}
		{Scientific Reports}\ }\textbf {\bibinfo {volume} {4}},\ \bibinfo {pages}
	{5886} (\bibinfo {year} {2014})}\BibitemShut {NoStop}%
\bibitem [{\citenamefont {Barth}, \citenamefont {Burger},\ and\ \citenamefont
	{Becker}(2016)}]{Barth2016}%
\BibitemOpen
\bibfield  {author} {\bibinfo {author} {\bibfnamefont {C.}~\bibnamefont
		{Barth}}, \bibinfo {author} {\bibfnamefont {S.}~\bibnamefont {Burger}}, \
	and\ \bibinfo {author} {\bibfnamefont {C.}~\bibnamefont {Becker}},\
}\bibfield  {title} {\enquote {\bibinfo {title} {{Symmetry-dependency of
			anticrossing phenomena in slab-type photonic crystals}},}\ }\href {\doibase
10.1364/OE.24.010931} {\bibfield  {journal} {\bibinfo  {journal} {Optics
		Express}\ }\textbf {\bibinfo {volume} {24}},\ \bibinfo {pages} {10931}
(\bibinfo {year} {2016})}\BibitemShut {NoStop}%
\bibitem [{\citenamefont {Sakoda}(2005)}]{Sakoda2004}%
\BibitemOpen
\bibfield  {author} {\bibinfo {author} {\bibfnamefont {K.}~\bibnamefont
		{Sakoda}},\ }\href {\doibase 10.1007/b138376} {\emph {\bibinfo {title}
		{{Optical Properties of Photonic Crystals}}}},\ \bibinfo {series} {Springer
	Series in Optical Sciences}, Vol.~\bibinfo {volume} {80}\ (\bibinfo
{publisher} {Springer-Verlag},\ \bibinfo {address} {Berlin/Heidelberg},\
\bibinfo {year} {2005})\ p.\ \bibinfo {pages} {258}\BibitemShut {NoStop}%
\bibitem [{\citenamefont {Rousseeuw}(1987)}]{Rousseeuw1987}%
\BibitemOpen
\bibfield  {author} {\bibinfo {author} {\bibfnamefont {P.~J.}\ \bibnamefont
		{Rousseeuw}},\ }\bibfield  {title} {\enquote {\bibinfo {title} {{Silhouettes:
				A graphical aid to the interpretation and validation of cluster analysis}},}\
}\href {\doibase 10.1016/0377-0427(87)90125-7} {\bibfield  {journal}
{\bibinfo  {journal} {Journal of Computational and Applied Mathematics}\
}\textbf {\bibinfo {volume} {20}},\ \bibinfo {pages} {53--65} (\bibinfo
{year} {1987})},\ \Eprint {http://arxiv.org/abs/z0024} {arXiv:z0024}
\BibitemShut {NoStop}%
\bibitem [{\citenamefont {Carey}\ \emph {et~al.}(2015)\citenamefont {Carey},
	\citenamefont {Abdelhady}, \citenamefont {Ning}, \citenamefont {Thon},
	\citenamefont {Bakr},\ and\ \citenamefont {Sargent}}]{Carey2015a}%
\BibitemOpen
\bibfield  {author} {\bibinfo {author} {\bibfnamefont {G.~H.}\ \bibnamefont
		{Carey}}, \bibinfo {author} {\bibfnamefont {A.~L.}\ \bibnamefont
		{Abdelhady}}, \bibinfo {author} {\bibfnamefont {Z.}~\bibnamefont {Ning}},
	\bibinfo {author} {\bibfnamefont {S.~M.}\ \bibnamefont {Thon}}, \bibinfo
	{author} {\bibfnamefont {O.~M.}\ \bibnamefont {Bakr}}, \ and\ \bibinfo
	{author} {\bibfnamefont {E.~H.}\ \bibnamefont {Sargent}},\ }\bibfield
{title} {\enquote {\bibinfo {title} {{Colloidal Quantum Dot Solar Cells}},}\
}\href {\doibase 10.1021/acs.chemrev.5b00063} {\bibfield  {journal} {\bibinfo
	{journal} {Chemical Reviews}\ }\textbf {\bibinfo {volume} {115}},\ \bibinfo
{pages} {12732--12763} (\bibinfo {year} {2015})}\BibitemShut {NoStop}%
\bibitem [{\citenamefont {Kamat}(2008)}]{Kamat2008a}%
\BibitemOpen
\bibfield  {author} {\bibinfo {author} {\bibfnamefont {P.~V.}\ \bibnamefont
		{Kamat}},\ }\bibfield  {title} {\enquote {\bibinfo {title} {{Quantum Dot
				Solar Cells. Semiconductor Nanocrystals as Light Harvesters}},}\ }\href
{\doibase 10.1021/jp806791s} {\bibfield  {journal} {\bibinfo  {journal} {The
			Journal of Physical Chemistry C}\ }\textbf {\bibinfo {volume} {112}},\
	\bibinfo {pages} {18737--18753} (\bibinfo {year} {2008})}\BibitemShut
{NoStop}%
\bibitem [{\citenamefont {Kojima}\ \emph {et~al.}(2009)\citenamefont {Kojima},
	\citenamefont {Teshima}, \citenamefont {Shirai},\ and\ \citenamefont
	{Miyasaka}}]{AkihiroKojimaKenjiroTeshimaYasuoShirai2009}%
\BibitemOpen
\bibfield  {author} {\bibinfo {author} {\bibfnamefont {A.}~\bibnamefont
		{Kojima}}, \bibinfo {author} {\bibfnamefont {K.}~\bibnamefont {Teshima}},
	\bibinfo {author} {\bibfnamefont {Y.}~\bibnamefont {Shirai}}, \ and\ \bibinfo
	{author} {\bibfnamefont {T.}~\bibnamefont {Miyasaka}},\ }\bibfield  {title}
{\enquote {\bibinfo {title} {{Organometal Halide Perovskites as Visible-Light
				Sensitizers for Photovoltaic Cells}},}\ }\href {\doibase 10.1021/ja809598r}
{\bibfield  {journal} {\bibinfo  {journal} {Journal of the American Chemical
			Society}\ }\textbf {\bibinfo {volume} {131}},\ \bibinfo {pages} {6050--6051}
	(\bibinfo {year} {2009})},\ \Eprint {http://arxiv.org/abs/nn504795v}
{arXiv:nn504795v [10.1021]} \BibitemShut {NoStop}%
\bibitem [{\citenamefont {Kundu}\ and\ \citenamefont
	{Patra}(2017)}]{Kundu2017}%
\BibitemOpen
\bibfield  {author} {\bibinfo {author} {\bibfnamefont {S.}~\bibnamefont
		{Kundu}}\ and\ \bibinfo {author} {\bibfnamefont {A.}~\bibnamefont {Patra}},\
}\bibfield  {title} {\enquote {\bibinfo {title} {{Nanoscale Strategies for
			Light Harvesting}},}\ }\href {\doibase 10.1021/acs.chemrev.6b00036}
{\bibfield  {journal} {\bibinfo  {journal} {Chemical Reviews}\ }\textbf
	{\bibinfo {volume} {117}},\ \bibinfo {pages} {712--757} (\bibinfo {year}
	{2017})}\BibitemShut {NoStop}%
\bibitem [{\citenamefont {Wu}\ \emph {et~al.}(2015)\citenamefont {Wu},
	\citenamefont {Congreve}, \citenamefont {Wilson}, \citenamefont {Jean},
	\citenamefont {Geva}, \citenamefont {Welborn}, \citenamefont {{Van Voorhis}},
	\citenamefont {Bulovi{\'{c}}}, \citenamefont {Bawendi},\ and\ \citenamefont
	{Baldo}}]{Wu2015b}%
\BibitemOpen
\bibfield  {author} {\bibinfo {author} {\bibfnamefont {M.}~\bibnamefont
		{Wu}}, \bibinfo {author} {\bibfnamefont {D.~N.}\ \bibnamefont {Congreve}},
	\bibinfo {author} {\bibfnamefont {M.~W.~B.}\ \bibnamefont {Wilson}}, \bibinfo
	{author} {\bibfnamefont {J.}~\bibnamefont {Jean}}, \bibinfo {author}
	{\bibfnamefont {N.}~\bibnamefont {Geva}}, \bibinfo {author} {\bibfnamefont
		{M.}~\bibnamefont {Welborn}}, \bibinfo {author} {\bibfnamefont
		{T.}~\bibnamefont {{Van Voorhis}}}, \bibinfo {author} {\bibfnamefont
		{V.}~\bibnamefont {Bulovi{\'{c}}}}, \bibinfo {author} {\bibfnamefont {M.~G.}\
		\bibnamefont {Bawendi}}, \ and\ \bibinfo {author} {\bibfnamefont {M.~A.}\
		\bibnamefont {Baldo}},\ }\bibfield  {title} {\enquote {\bibinfo {title}
		{{Solid-state infrared-to-visible upconversion sensitized by colloidal
				nanocrystals}},}\ }\href {\doibase 10.1038/nphoton.2015.226} {\bibfield
	{journal} {\bibinfo  {journal} {Nature Photonics}\ }\textbf {\bibinfo
		{volume} {10}},\ \bibinfo {pages} {31--34} (\bibinfo {year}
	{2015})}\BibitemShut {NoStop}%
\bibitem [{\citenamefont {Wu}, \citenamefont {Congreve},\ and\ \citenamefont
	{Baldo}(2015)}]{Wu2015c}%
\BibitemOpen
\bibfield  {author} {\bibinfo {author} {\bibfnamefont {T.~C.}\ \bibnamefont
		{Wu}}, \bibinfo {author} {\bibfnamefont {D.~N.}\ \bibnamefont {Congreve}}, \
	and\ \bibinfo {author} {\bibfnamefont {M.~A.}\ \bibnamefont {Baldo}},\
}\bibfield  {title} {\enquote {\bibinfo {title} {{Solid state photon
			upconversion utilizing thermally activated delayed fluorescence molecules as
			triplet sensitizer}},}\ }\href {\doibase 10.1063/1.4926914} {\bibfield
{journal} {\bibinfo  {journal} {Applied Physics Letters}\ }\textbf {\bibinfo
	{volume} {107}} (\bibinfo {year} {2015}),\ 10.1063/1.4926914}\BibitemShut
{NoStop}%
\bibitem [{\citenamefont {Pedregosa}\ \emph {et~al.}(2012)\citenamefont
	{Pedregosa}, \citenamefont {Varoquaux}, \citenamefont {Gramfort},
	\citenamefont {Michel}, \citenamefont {Thirion}, \citenamefont {Grisel},
	\citenamefont {Blondel}, \citenamefont {Louppe}, \citenamefont
	{Prettenhofer}, \citenamefont {Weiss}, \citenamefont {Dubourg}, \citenamefont
	{Vanderplas}, \citenamefont {Passos}, \citenamefont {Cournapeau},
	\citenamefont {Brucher}, \citenamefont {Perrot},\ and\ \citenamefont
	{Duchesnay}}]{Pedregosa2012}%
\BibitemOpen
\bibfield  {author} {\bibinfo {author} {\bibfnamefont {F.}~\bibnamefont
		{Pedregosa}}, \bibinfo {author} {\bibfnamefont {G.}~\bibnamefont
		{Varoquaux}}, \bibinfo {author} {\bibfnamefont {A.}~\bibnamefont {Gramfort}},
	\bibinfo {author} {\bibfnamefont {V.}~\bibnamefont {Michel}}, \bibinfo
	{author} {\bibfnamefont {B.}~\bibnamefont {Thirion}}, \bibinfo {author}
	{\bibfnamefont {O.}~\bibnamefont {Grisel}}, \bibinfo {author} {\bibfnamefont
		{M.}~\bibnamefont {Blondel}}, \bibinfo {author} {\bibfnamefont
		{G.}~\bibnamefont {Louppe}}, \bibinfo {author} {\bibfnamefont
		{P.}~\bibnamefont {Prettenhofer}}, \bibinfo {author} {\bibfnamefont
		{R.}~\bibnamefont {Weiss}}, \bibinfo {author} {\bibfnamefont
		{V.}~\bibnamefont {Dubourg}}, \bibinfo {author} {\bibfnamefont
		{J.}~\bibnamefont {Vanderplas}}, \bibinfo {author} {\bibfnamefont
		{A.}~\bibnamefont {Passos}}, \bibinfo {author} {\bibfnamefont
		{D.}~\bibnamefont {Cournapeau}}, \bibinfo {author} {\bibfnamefont
		{M.}~\bibnamefont {Brucher}}, \bibinfo {author} {\bibfnamefont
		{M.}~\bibnamefont {Perrot}}, \ and\ \bibinfo {author} {\bibfnamefont
		{{\'{E}}.}~\bibnamefont {Duchesnay}},\ }\bibfield  {title} {\enquote
	{\bibinfo {title} {{Scikit-learn: Machine Learning in Python}},}\ }\href
{\doibase 10.1007/s13398-014-0173-7.2} {\ \textbf {\bibinfo {volume} {12}},\
	\bibinfo {pages} {2825--2830} (\bibinfo {year} {2012})},\ \Eprint
{http://arxiv.org/abs/1201.0490} {arXiv:1201.0490} \BibitemShut {NoStop}%
\bibitem [{\citenamefont {Software}(2013)}]{POVRay2013}%
\BibitemOpen
\bibfield  {author} {\bibinfo {author} {\bibnamefont {Software}},\ }\bibfield
{title} {\enquote {\bibinfo {title} {{Persistence of Vision (TM) Raytracer
				(POV-Ray), Version 3.7}},}\ }\href@noop {} {\bibfield  {journal} {\bibinfo
		{journal} {Persistence of Vision Pty. Ltd.}\ } (\bibinfo {year}
	{2013})}\BibitemShut {NoStop}%
\bibitem [{\citenamefont {Schulze}\ and\ \citenamefont
	{Schmidt}(2015)}]{Schulze2014a}%
\BibitemOpen
\bibfield  {author} {\bibinfo {author} {\bibfnamefont {T.~F.}\ \bibnamefont
		{Schulze}}\ and\ \bibinfo {author} {\bibfnamefont {T.~W.}\ \bibnamefont
		{Schmidt}},\ }\bibfield  {title} {\enquote {\bibinfo {title} {{Photochemical
				upconversion: present status and prospects for its application to solar
				energy conversion}},}\ }\href {\doibase 10.1039/C4EE02481H} {\bibfield
	{journal} {\bibinfo  {journal} {Energy {\&} Environmental Science}\ }\textbf
	{\bibinfo {volume} {8}},\ \bibinfo {pages} {103--125} (\bibinfo {year}
	{2015})},\ \Eprint {http://arxiv.org/abs/{\_}barata Materials and Techniques
	of polychrome wooden sculpture} {arXiv:{\_}barata Materials and Techniques of
	polychrome wooden sculpture} \BibitemShut {NoStop}%
\bibitem [{\citenamefont {Park}, \citenamefont {Lu},\ and\ \citenamefont
	{Ahn}(2015)}]{Park2015}%
\BibitemOpen
\bibfield  {author} {\bibinfo {author} {\bibfnamefont {W.}~\bibnamefont
		{Park}}, \bibinfo {author} {\bibfnamefont {D.}~\bibnamefont {Lu}}, \ and\
	\bibinfo {author} {\bibfnamefont {S.}~\bibnamefont {Ahn}},\ }\bibfield
{title} {\enquote {\bibinfo {title} {{Plasmon enhancement of luminescence
				upconversion}},}\ }\href {\doibase 10.1039/C5CS00050E} {\bibfield  {journal}
	{\bibinfo  {journal} {Chem. Soc. Rev.}\ }\textbf {\bibinfo {volume} {44}},\
	\bibinfo {pages} {2940--2962} (\bibinfo {year} {2015})}\BibitemShut {NoStop}%
\bibitem [{\citenamefont {Bishop}(2006)}]{Bishop2013}%
\BibitemOpen
\bibfield  {author} {\bibinfo {author} {\bibfnamefont {C.~M.}\ \bibnamefont
		{Bishop}},\ }\href
{http://electronicimaging.spiedigitallibrary.org/article.aspx?doi=10.1117/1.2819119
	http://www.ncbi.nlm.nih.gov/pubmed/25246403
	http://www.pubmedcentral.nih.gov/articlerender.fcgi?artid=PMC4249520
	http://arxiv.org/abs/1011.1669 http://dx.doi.org/10.1088/17} {\emph {\bibinfo
		{title} {{Pattern Recognition and Machine Learning}}}}\ (\bibinfo
{publisher} {Springer US},\ \bibinfo {year} {2006})\BibitemShut {NoStop}%
\bibitem [{\citenamefont {Hastie}, \citenamefont {Tibshirani},\ and\
	\citenamefont {Friedman}(2009)}]{Mining2009}%
\BibitemOpen
\bibfield  {author} {\bibinfo {author} {\bibfnamefont {T.}~\bibnamefont
		{Hastie}}, \bibinfo {author} {\bibfnamefont {R.}~\bibnamefont {Tibshirani}},
	\ and\ \bibinfo {author} {\bibfnamefont {J.}~\bibnamefont {Friedman}},\
}\href {\doibase 10.1007/978-0-387-84858-7} {\emph {\bibinfo {title} {{The
			Elements of Statistical Learning}}}},\ \bibinfo {edition} {2nd}\ ed.,\
\bibinfo {series} {Springer Series in Statistics}, Vol.~\bibinfo {volume}
{27}\ (\bibinfo  {publisher} {Springer New York},\ \bibinfo {address} {New
	York, NY},\ \bibinfo {year} {2009})\ pp.\ \bibinfo {pages}
{83--85}\BibitemShut {NoStop}%
\bibitem [{\citenamefont {Dempster}, \citenamefont {Laird},\ and\ \citenamefont
	{Rubin}(1977)}]{Dempster1977}%
\BibitemOpen
\bibfield  {author} {\bibinfo {author} {\bibfnamefont {A.~P.}\ \bibnamefont
		{Dempster}}, \bibinfo {author} {\bibfnamefont {N.~M.}\ \bibnamefont {Laird}},
	\ and\ \bibinfo {author} {\bibfnamefont {D.~B.}\ \bibnamefont {Rubin}},\
}\bibfield  {title} {\enquote {\bibinfo {title} {{Maximum Likelihood from
			Incomplete Data via the EM Algorithm}},}\ }\href {\doibase 10.2307/2984875}
{\bibfield  {journal} {\bibinfo  {journal} {Journal of the Royal Statistical
			Society Series B (Methodological)}\ }\textbf {\bibinfo {volume} {39}},\
	\bibinfo {pages} {1--38} (\bibinfo {year} {1977})},\ \Eprint
{http://arxiv.org/abs/0710.5696v2} {arXiv:0710.5696v2} \BibitemShut {NoStop}%
\bibitem [{\citenamefont {McLachlan}\ and\ \citenamefont
	{Krishnan}(1997)}]{McLachlan1997}%
\BibitemOpen
\bibfield  {author} {\bibinfo {author} {\bibfnamefont {G.~J.}\ \bibnamefont
		{McLachlan}}\ and\ \bibinfo {author} {\bibfnamefont {T.}~\bibnamefont
		{Krishnan}},\ }\href {\doibase 10.1002/9780470191613.scard} {\emph {\bibinfo
		{title} {The EM Algorithm and Extensions}}},\ \bibinfo {edition} {2nd}\ ed.\
(\bibinfo  {publisher} {John Wiley {\&} Sons, Inc.},\ \bibinfo {address}
{Hoboken, NJ, USA},\ \bibinfo {year} {1997})\BibitemShut {NoStop}%
\end{thebibliography}
\end{document}